\documentclass[aps,prb,twocolumn,showpacs]{revtex4}
\usepackage{graphicx}
\usepackage{natbib}

\begin{document}
\title{Temperature dependence of spatially heterogeneous dynamics in a model of viscous silica}

\author{Michael Vogel}
\altaffiliation{Institut f\"ur Physikalische Chemie,
Westf\"alische Wilhelms-Universit\"at M\"unster, Corensstr. 30,
48149 M\"unster, Germany} \email{mivogel@uni-muenster.de}
\author{Sharon C. Glotzer}
\email{sglotzer@umich.edu} \affiliation{Department of Chemical
Engineering and Department of Materials Science and Engineering,
University of Michigan, 2300 Hayward, Ann Arbor, MI, 48109, USA}

\date{\today}

\begin{abstract}
Molecular dynamics simulations are performed to study spatially
heterogeneous dynamics in a model of viscous silica above and
below the critical temperature of the mode coupling theory,
$T_{MCT}$. Specifically, we follow the evolution of the dynamic
heterogeneity as the temperature dependence of the transport
coefficients shows a crossover from non-Arrhenius to Arrhenius
behavior when the melt is cooled. It is demonstrated that, on
intermediate time scales, a small fraction of oxygen and silicon
atoms are more mobile than expected from a Gaussian approximation.
These highly mobile particles form transient clusters larger than
that resulting from random statistics, indicating that dynamics
are spatially heterogeneous. An analysis of the clusters reveals
that the mean cluster size is maximum at times intermediate
between ballistic and diffusive motion, and the maximum size
increases with decreasing temperature. In particular, the growth
of the clusters continues when the transport coefficients follow
an Arrhenius law. These findings imply that the structural
relaxation in silica cannot be understood as a statistical bond
breaking process. Though the mean cluster sizes for silica are at
the lower end of the spectrum of values reported in the
literature, we find that spatially heterogeneous dynamics in
strong and fragile glass formers are similar on a qualitative
level. However, different from results for fragile liquids, we
show that correlated particle motion along quasi one-dimensional,
string-like paths is of little importance for the structural
relaxation in this model of silica, suggesting that string-like
motion is suppressed by the presence of covalent bonds. To study
spatial correlations between highly immobile particles, we
calculate a generalized susceptibility corresponding to the self
part of a four-point time dependent density correlation function.
We find that this generalized susceptibility is maximum on the
time scale of the structural relaxation, where a strong increase
of the peak height indicates a growing length of spatial
correlations between highly immobile particles upon cooling.
Characterizing the local structures of the most mobile and the
most immobile particles, respectively, we show that high particle
mobility is facilitated by, but not limited to, the vicinity of
defects of the network structure.

\end{abstract}

 \maketitle

\section{Introduction}

Applications of various experimental methods demonstrated that the
structural relaxation of many supercooled liquids exhibits dynamic
heterogeneities, i.e., it is possible to select a subset of
particles that rotate or translate much farther or shorter
distances than an average
particle.~\cite{HWS,ME-PB,12,HS,ME-RE,VDB} However, the majority
of experimental techniques do not provide information about the
spatial arrangement of mobile and immobile particles, which plays
a central role in several theories of the glass
transition.~\cite{AG,GC-PRL,GC-PN} Recently, results of
multidimensional nuclear magnetic resonance experiments showed
that dynamics in polymers~\cite{AH-4D} and in organic
low-molecular weight compounds~\cite{ME-4D,ME-4D2} are spatially
heterogeneous. Particles within a physical region of these viscous
liquids show an enhanced or diminished mobility compared to
particles in a region a few nanometers away. Despite this
progress, an experimental characterization of the time evolution
of spatially heterogeneous dynamics (SHD) is still lacking.
Further, detailed experimental studies of the dynamical behavior
of strong liquids, like silica, are hampered by the high glass
transition temperatures, $T_g$, of these materials. Nevertheless,
based on the Arrhenius behavior observed at sufficiently low
temperatures, it is often argued that the dynamics in silica melts
can be understood as a statistical bond breaking process.

Molecular dynamics (MD) simulations of model glass-forming liquids
provide direct access to spatial correlations of particle mobility.
Due to the available computer power, present simulations are
performed at temperatures near $T_{MCT}$, the critical temperature of
the mode-coupling theory (MCT),~\cite{MCT} where the structural
relaxation typically occurs on the order of nanoseconds. In previous
work, MD simulations were used to investigate dynamic heterogeneity
in soft-sphere~\cite{DP,YH,AO} and hard-sphere systems,~\cite{BD-HS}
binary Lennard-Jones (BLJ)
mixtures,~\cite{GW,GC,WK,CD-PRL,CD-PRE,CD-PRL2,SCG-JCP,NL,NL2,LB} the
Dzugutov liquid,~\cite{MD,YG-DZ1,YG-DZ2} polymer
melts~\cite{CB,YG-PO,MA,AH-PO} and molecular liquids.~\cite{AH-PC,NG}
These studies showed that highly mobile and highly immobile particles
aggregate into clusters that are transient in nature.

Specifically, highly mobile particles both in simple liquids, such
as BLJ mixtures~\cite{CD-PRE} and the Dzugutov
liquid,~\cite{YG-DZ2} and in more complex systems, like polymer
melts~\cite{YG-PO} and water,~\cite{NG} form clusters that are
largest at times intermediate between ballistic and diffusive
motion. The size of these clusters grows strongly when the liquid
is cooled and a divergence at $T\!\approx\!T_{MCT}$ was
proposed.~\cite{CD-PRE,SCG} Within the clusters smaller groups of
particles move in strings, i.e., particles follow each other along
quasi one-dimensional paths. This correlated motion on
intermediate time scales, which is believed to facilitate
structural relaxation,~\cite{SCG} appears to be universal as well,
as it is found for a range of different model
liquids.~\cite{CD-PRL,CD-PRE,YG-DZ2,MA} In addition, spatial
correlations of highly immobile particles were studied based on a
four-point time dependent density correlation
function.~\cite{SCG-JCP,NL,NL2,LB,SF,CDA} For BLJ liquids, it was
observed that the corresponding generalized susceptibility
$\chi_4(\Delta t)$ exhibits a maximum that increases with
decreasing temperature, indicating that there is a growing length
of spatial correlations between particles localized in a time
interval $\Delta t$.

All these studies were performed in temperature ranges where the
evolution of the relaxation times shows pronounced deviations from
Arrhenius behavior. In a recent MD simulation study,~\cite{MV-PRL} we
demonstrated that SHD exists not only in fragile liquids, but also in
a model of silica, the paradigm of a strong liquid.~\cite{AA}
Specifically, the most mobile particles were found to form clusters
larger than predicted from random statistics, where the mean cluster
size is maximum at times intermediate between ballistic and diffusive
motion, as was observed for other model liquids. Further, we showed
that the dynamic heterogeneities are short lived so that high
particle mobility spreads throughout the sample on the time scale of
the structural relaxation. In doing so, the probability for a
previously immobile particle to become mobile is enhanced in the
vicinity of another mobile particle, supporting the concept of
dynamic facilitation, which is the cornerstone of a microscopic model
of viscous liquids proposed recently by Garrahan and
Chandler.~\cite{GC-PRL,GC-PN} Finally, our previous results suggest
that, compared to other model liquids, string-like motion is less
relevant for the structural relaxation of silica.

In view of these findings for fragile and strong liquids, one may
ask how the transport coefficients of viscous liquids -- in
particular their temperature dependence --  are related to the
properties of SHD. Indeed, the decoupling of diffusivity and
viscosity or structural relaxation time, indicating a breakdown of
the classic Stokes-Einstein relation, has been rationalized in
terms of SHD.~\cite{ME-RE,SCG-JCP} Furthermore, a link between the
diffusion coefficient $D$ and the clusters of mobile particles was
observed in simulation studies of water.~\cite{NG} Specifically,
it was reported that the mean cluster size is a measure of the
mass of the cooperatively rearranging regions central to the
Adam-Gibbs theory,~\cite{AG} which in turn allows one to describe
the temperature dependence of $D$. This suggests that the behavior
of the transport coefficients, which describe the long-time liquid
dynamics, is determined by SHD at intermediate times. On the other
hand, Garrahan and Chandler~\cite{GC-PRL,GC-PN} emphasize that
spatiotemporal characteristics of mobility propagation are
important to rationalize differences between various glass
formers. In particular, they argue that mobility propagation
carries a direction, the persistence length of which is larger for
fragile than for strong glass formers.

Here, we continue our MD simulation study of SHD in the BKS model
of viscous silica,~\cite{BKS} which is commonly used to reproduce
structural and dynamical properties of this strong
liquid.~\cite{JB,JH-ALL,JH-FS,SV,WK-REV,KB,KV,JH-BP1,JH-BP2} In
this way, we wish to obtain further insights into the relaxation
dynamics of one of the most important glass formers, which are
difficult to extract from experimental studies. It is well
established that the transport coefficients of BKS silica show a
crossover from a non-Arrhenius to an Arrhenius temperature
dependence upon cooling,~\cite{JH-ALL,JH-FS,SV,WK-REV,KB} in
agreement with experimental data.~\cite{ER} Saika-Voivod et
al.~\cite{SV} related such behavior to a ``fragile-to-strong
transition''. A main goal of the present work is to quantify the
evolution of SHD during this crossover. Moreover, by means of
quantitative comparison with literature data, we seek to ascertain
differences and similarities of SHD among fragile and strong
liquids, respectively. We study spatial correlations both between
highly mobile particles and between highly immobile particles.
Finally, we analyze the relation between properties of the local
instantaneous liquid structure and the particle mobility, to
investigate whether there is a structural origin of SHD in silica.
Indications of a relationship between local structure and local
dynamics have been demonstrated in the Dzugutov~\cite{MD,MB} and
BLJ liquids.~\cite{CD-PRE}

\section{Model and Simulation}\label{Model}

Previous MD simulation
studies~\cite{JB,JH-ALL,JH-FS,SV,WK-REV,KB,KV,JH-BP1,JH-BP2} have
shown that the BKS potential~\cite{BKS} is well suited to
reproduce many structural and dynamical properties of amorphous
silica. The BKS potential energy is given by
\begin{equation}
\phi_{\alpha\beta}(r)=\frac{q_{\alpha}q_{\beta}e^2}{r}+A_{\alpha\beta}\exp(-B_{\alpha\beta}r)-\frac{C_{\alpha\beta}}{r^6}
\end{equation}
where $r$ is the distance between two atoms of types $\alpha$ and
$\beta$ ($\alpha,\,\beta\in\{\mathrm{Si,O}\}$) and $e$ denotes the
elementary charge. The partial charges $q_{\alpha}$ together with the
potential parameters $A_{\alpha\beta}$, $B_{\alpha\beta}$ and
$C_{\alpha\beta}$ can be found elsewhere.~\cite{BKS,KV}

Here, we follow simulations of Horbach and Kob~\cite{JH-ALL,JH-FS}
and perform computations in the NVE ensemble where $N\!=\!8016$
and $\rho\!=\!2.37\,\mathrm{g/cm^3}$ corresponding to a box length
$L\!=\!48.4\,\mathrm \AA$. This system size is sufficiently large
so as to avoid significant finite size effects at the studied
$T$.~\cite{JH-BP1,JH-BP2} The equations of motion are integrated
using the velocity Verlet algorithm with a time step of
$1.6\,\mathrm{fs}$. In doing so, we truncate the non-Coulombic
part of the potential at a cutoff radius of $5.5\,\mathrm \AA$.
The Coulombic part is calculated via Ewald summation~\cite{AT}
where we apply a parameter $\alpha\!=\!0.265$ together with a
cutoff in Fourier space $k_c\!=\!2\pi/L\sqrt{51}$. As was
discussed in previous work,~\cite{JH-ALL,JH-FS} the use of this
value of $k_c$ leads to a constant shift of the total potential
energy and, hence, does not affect the forces. We consider values
of $T$ in a range between $3030\,\mathrm{K}$ and
$5250\,\mathrm{K}$, where the system is equilibrated for times
longer than the structural relaxation time prior to data
accumulation.

Horbach and Kob~\cite{JH-ALL,JH-FS} showed that BKS silica
consists of a network of well defined silicate tetrahedra.
Further, they analyzed the temperature dependences of the
diffusion coefficient $D$ and the time constant $\tau$ of the
structural relaxation. At relatively high $T$, these quantities
were found to vary according to a power-law
\begin{equation}\label{PL}
D,\;\tau^{-1}\propto(T-T_{MCT})^{\gamma},
\end{equation}
as predicted by MCT.~\cite{MCT} For the critical temperature,
Horbach and Kob~\cite{JH-ALL,JH-FS} reported
$T_{MCT}\!=\!3330\,\mathrm K$, in good agreement with
$T_{MCT}\!=\!3221\,\mathrm K$ determined from experimental
viscosity data of silica.~\cite{ER} The critical exponents
extracted from $D(T)$ and $\tau^{-1}(T)$ were found to be
$\gamma\!\approx\!2.1$ and $\gamma\!\approx\!2.4$,
respectively.~\cite{JH-ALL,JH-FS} At $T$ below a dynamical
crossover in the vicinity of $T_{MCT}$, Arrhenius laws
\begin{equation}\label{ARR}
D,\;\tau^{-1}\propto\exp\left[-E_a/(k_B T)\right]
\end{equation}
were observed.~\cite{JH-ALL,JH-FS} Activation energies
$E_a\!=\!4.7\,\mathrm{eV}$ and $E_a\!=\!5.2\,\mathrm{eV}$ for the
oxygen and silicon atoms, respectively, were obtained from $D(T)$.
The temperature dependence $\tau^{-1}(T)$, extracted from the
incoherent intermediate scattering functions for the oxygen atoms,
yielded activation energies between $5.0\,\mathrm{eV}$ and
$5.5\,\mathrm{eV}$, depending on the wave vector $q$. All these
values are similar to $E_a\!=\!4.7\,\mathrm{eV}$ for oxygen and
$E_a\!=\!6.0\,\mathrm{eV}$ for silicon observed in experimental
studies near $T_g$.~\cite{EAO,EAS}

Our analysis confirms these findings of Horbach and
Kob.~\cite{JH-ALL,JH-FS} Specifically, our data are consistent with
$T_{MCT}\!=\!3330\,\mathrm K$ and we obtain critical exponents
$\gamma\!\approx\!2.0$ and $\gamma\!\approx\!2.2$ from $D(T)$ and
$\tau^{-1}(T)$, respectively. Furthermore, we observe comparable
activation energies, see section~\ref{Bulk}.

\section{Results}
\subsection{Properties of the bulk}\label{Bulk}

\begin{figure}
\includegraphics[angle=0,width=7.5cm,clip]{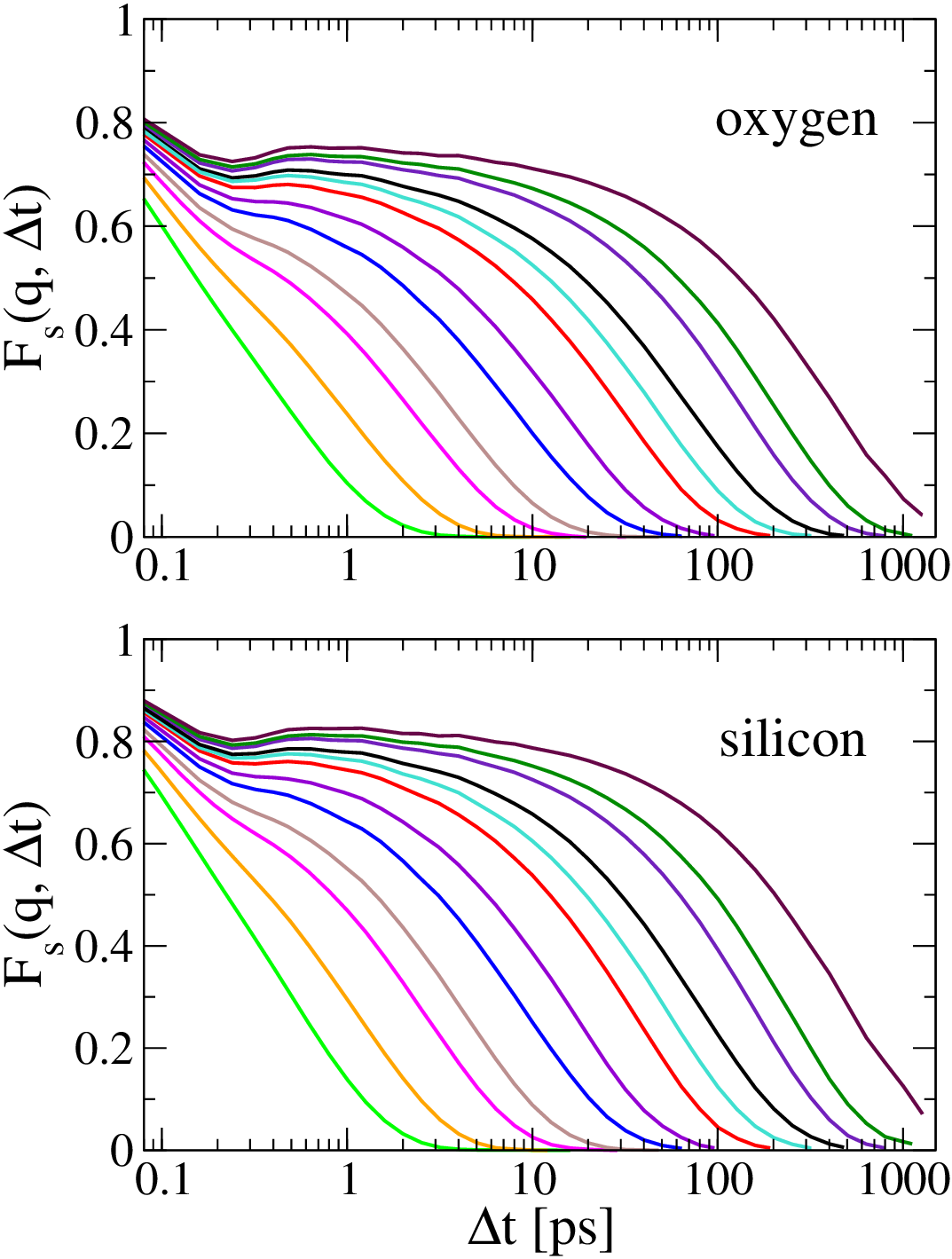}
\caption{Incoherent intermediate scattering functions,
$F_s(q\!=\!1.7\mathrm\AA^{-1};\Delta t)$, for the oxygen and
silicon atoms, respectively. The temperatures are, from left to
right: $5250\,\mathrm{K}$, $4730\,\mathrm{K}$, $4330\,\mathrm{K}$,
$4120\,\mathrm{K}$, $3870\,\mathrm{K}$, $3710\,\mathrm{K}$,
$3520\,\mathrm{K}$, $3420\,\mathrm{K}$, $3330\,\mathrm{K}$,
$3230\,\mathrm{K}$, $3140\,\mathrm{K}$ and $3030\,\mathrm{K}$.
}\label{FS}
\end{figure}

To mark the various time regimes of the structural relaxation in BKS
silica, we first discuss the incoherent intermediate scattering
function
\begin{equation}\label{ISF}
F_s(q,\Delta t)=\langle\,\cos\{\,\vec{q}\,[\vec{r}_{i}(t_0\!+\!\Delta
t)-\vec{r}_{i}(t_0)]\,\}\,\rangle.
\end{equation}
Here, $\vec{r}_i(t_0)$ is the position of particle $i$ at time
$t_0$ and the brackets $\langle \dots \rangle$ denote the ensemble
average. Figure~\ref{FS} shows $F_{s}(q,\Delta t)$ for the oxygen
and silicon atoms, respectively, where we use an absolute value of
the wave vector, $q\!=\!1.7\mathrm\AA^{-1}$. This value of $q$
corresponds to the first sharp diffraction peak of the static
structure factor~\cite{JH-ALL} and, hence, dynamics on the length
scale of the distance of the silicate tetrahedra is probed. We see
in Fig.~\ref{FS} that the scattering functions for both atomic
species are comparable. Typical of viscous liquids,
$F_{s}(q,\Delta t)$ shows a two-step decay. While the short-time
decay can be ascribed to ballistic motion, the non-exponential
long-time decay results from the structural relaxation. As a
consequence of the cage effect,~\cite{MCT} which describes that
the particles are temporarily trapped in a cage formed by their
neighbors, a plateau regime develops between the ballistic and
diffusive regimes upon cooling. At the crossover from the
ballistic to the plateau regime, the curves show oscillations that
are damped out for longer times. Horbach \emph{et
al}.~\cite{JH-BP1,JH-BP2} found a similar behavior and ascribed it
to the Boson peak. However, the origin of this feature is still a
matter of debate.~\cite{JH-FS,FS-H2O}

\begin{figure}
\includegraphics[angle=0,width=7.5cm,clip]{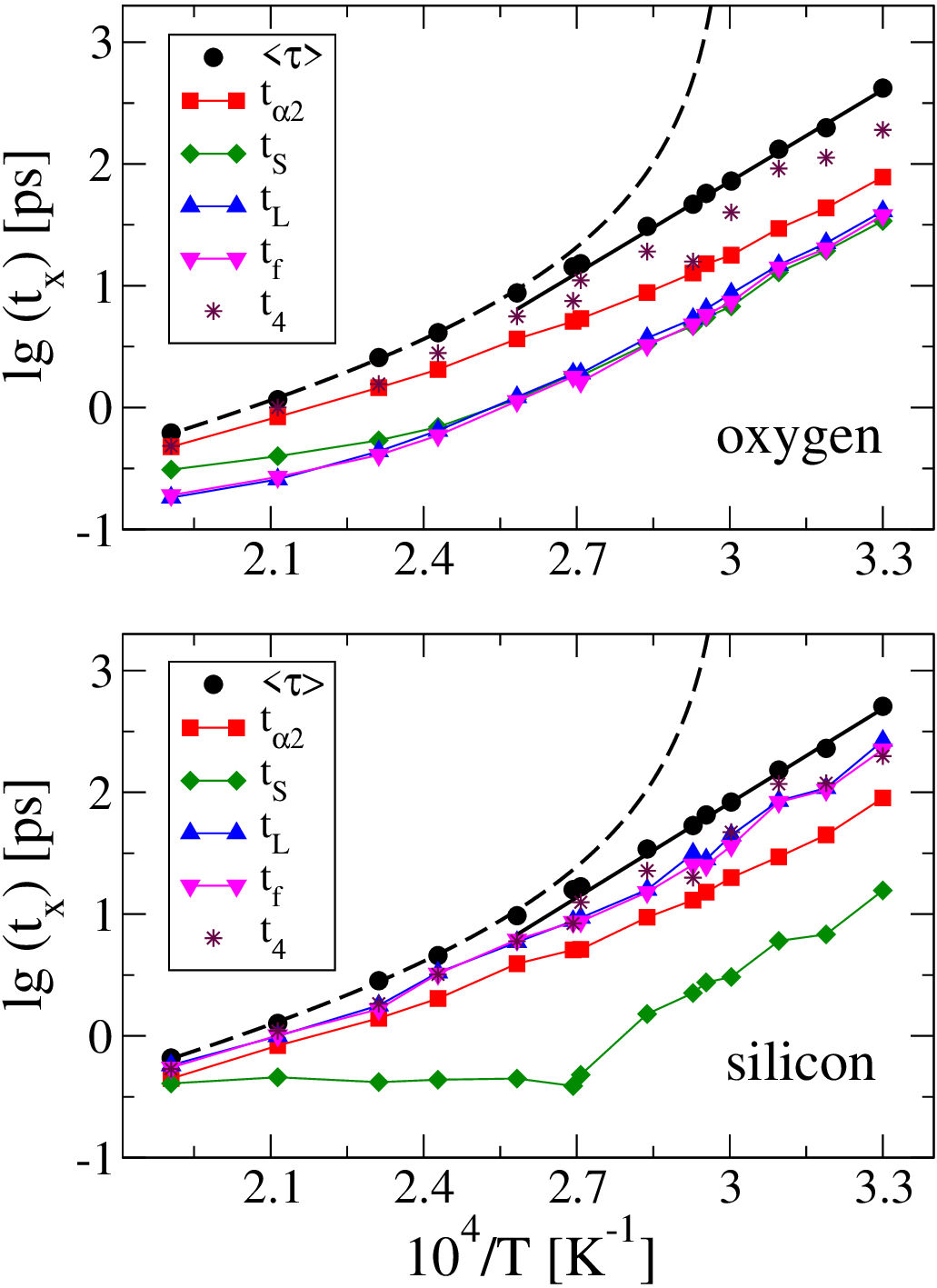}
\caption{Various time constants characterizing the dynamics of the
oxygen and silicon atoms in BKS silica, see text for details. The
data $\langle \tau\,(T) \rangle$ are fit by a power-law at
$T\!>\!3900\mathrm{\,K}$ (dashed lines, Eq.~\ref{PL} with
$T_{MCT}\!=\!3330 \mathrm{\,K}$, $\gamma\!=\!2.2$ for oxygen and
silicon) and by an Arrhenius law at $T\!<\!3300\mathrm{\,K}$ (solid
lines, Eq.~\ref{ARR}, $E_a\!=\!5.0\mathrm{\,eV}$ for oxygen,
$E_a\!=\!5.1\mathrm{\,eV}$ for silicon).}\label{TAU}
\end{figure}

For an analysis of the structural relaxation, we fit the long-time
decay of $F_s(q,\Delta t)$ to a Kohlrausch-Williams-Watts (KWW)
function, $A\,\exp[-(t/\tau)^\beta]$. In Fig.~\ref{TAU}, it is
evident that the temperature dependence of the mean time constant,
given by $\langle\tau \rangle=\!(\tau/\beta)\Gamma(1/\beta)$, is
similar for both atomic species. At
$T\!\lesssim\!3300\mathrm{\,K}$, the data are well described by an
Arrhenius law where we find activation energies
$E_a\!=\!5.0\mathrm{\,eV}$ and $E_a\!=\!5.1\mathrm{\,eV}$ for the
oxygen and silicon atoms, respectively. In contrast, there are
deviations from an Arrhenius law at higher $T$, reflecting the
crossover observed for the temperature dependence of the transport
coefficients.~\cite{JH-ALL,SV,MV-PRL} In particular, $\langle\tau
\,(T)\rangle$ can be fitted by Eq.~\ref{PL} with
$T_{MCT}\!=\!3330\,\mathrm K$ at $T\!>\!3900\mathrm{\,K}$. The
non-exponentiality of the structural relaxation can be quantified
by the stretching parameter $\beta$ of the KWW function. We
observe $\beta\!>\!0.8$ for all studied $T$ and for both atomic
species, and, hence, the deviations from an exponential behavior
are small. A closer inspection reveals that starting from
$\beta\!\approx\!1$ at high $T$ the stretching parameter first
decreases, and then becomes constant within statistical error at
$T\!\lesssim\!3800\mathrm{\,K}$ (oxygen:
$\beta\!=\!0.83\!\pm\!0.02$, silicon:
$\beta\!=\!0.85\!\pm\!0.02$).

\begin{figure}
\includegraphics[angle=0,width=7.5cm,clip]{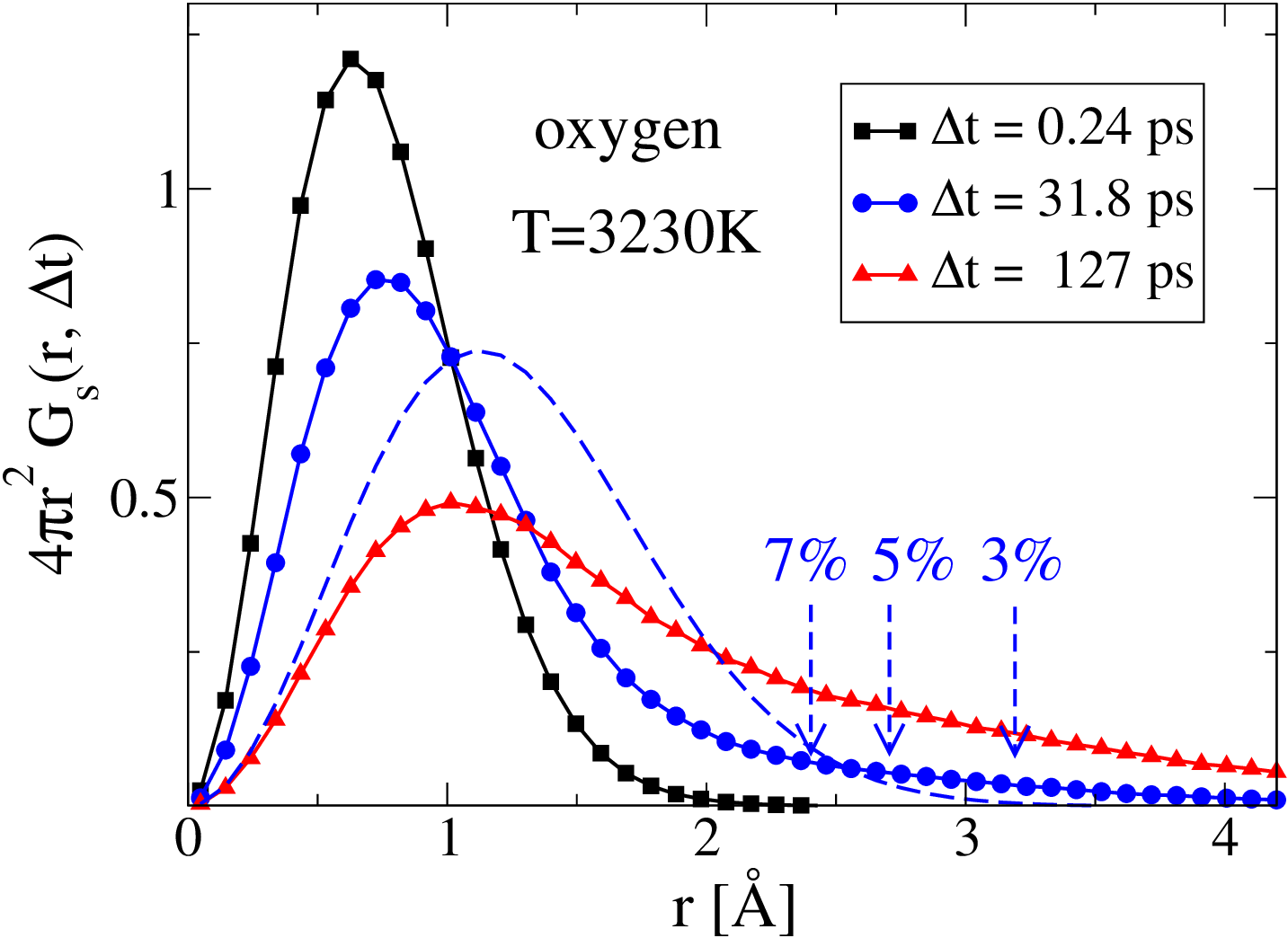}
\caption{Self part of the van Hove correlation function,
$G_s(r,\Delta t)$, for the oxygen atoms in BKS silica at
$T\!=\!3230\,\mathrm K$. The time $\Delta t\!=\!31.8\mathrm{\,ps}$
compares to the transition between the plateau and diffusive
regimes. Dashed line: Gaussian approximation $G_0(r,\Delta t)$ for
$\Delta t\!=\!31.8\mathrm{\,ps}$, cf.\ Eq.~\ref{Gauss}. The arrows
indicate the distances $r_c$ for which the integral
$\phi\!=\!\int_{r_c}^{\infty}4\pi r^2 G_s(r,\Delta
t\!=\!31.8\mathrm{ps}) dr$ equals 3\%, 5\% and 7\%, respectively.
The data to the right of the arrows result from highly mobile
particles studied in section~\ref{Mobile}. For comparison, the
oxygen-oxygen interparticle distance $r_{OO}$ equals
$2.6\mathrm{\,\AA}$, see Fig.~\ref{g(r)}}\label{VAN}.
\end{figure}

Next, we study the self part of the van Hove correlation function
\begin{equation}
G_s(\vec{r},\Delta t)=\langle\,
\delta\left[\,\vec{r}_i(t_0\!+\!\Delta t)-\vec{r}_i(t_0) - \vec{r}\,
\right]\,\rangle.
\end{equation}
The quantity $4\pi r^2\,G_s(r,\Delta t)$ measures the probability
that a particle moves a distance $r$ in a time interval $\Delta
t$. Figure~\ref{VAN} depicts this probability for the oxygen atoms
at $T\!=\!3230\,\mathrm K$. It is instructive to compare
$G_s(r,\Delta t)$ with the Gaussian approximation
\cite{WK,CD-PRE,YG-PO}
\begin{equation}\label{Gauss}
G_0 (r,\Delta t)=\left(\frac{3}{2\pi \langle r_i^2(\Delta t)\rangle
}\right)^{\frac{3}{2}}\;\exp\left(-\frac{3}{2\langle r_i^2(\Delta t)
\rangle }\right)
\end{equation}
where $r_i^2(\Delta t)\!=\!|\vec{r}_i(t_0\!+\!\Delta t)\!-
\!\vec{r}_i(t_0)|^2$. While $G_s(r,\Delta t)\!=\!G_0(r,\Delta t)$
for short and long times in the structural relaxation process,
significant deviations are obvious at intermediate times. They are
demonstrated for $\Delta t\!=\!31.8\mathrm{\,ps}$ in
Fig.~\ref{VAN}. For this time interval, $4\pi r^2\,G_s(r,\Delta
t)$ shows a pronounced tail, indicating that some particles move
much farther than expected from a Gaussian approximation. Such
behavior is well known from previous studies of fragile viscous
liquids.~\cite{WK,CD-PRE,YG-PO,MA,YG-DZ2,AH-PC,NG} In the present
system, we observe this behavior also at $T$ below the
fragile-to-strong crossover, consistent with previous
findings.~\cite{JH-ALL} This result gives a first indication of a
similarity in SHD between fragile and strong liquids.

\begin{figure}
\includegraphics[angle=0,width=7.5cm,clip]{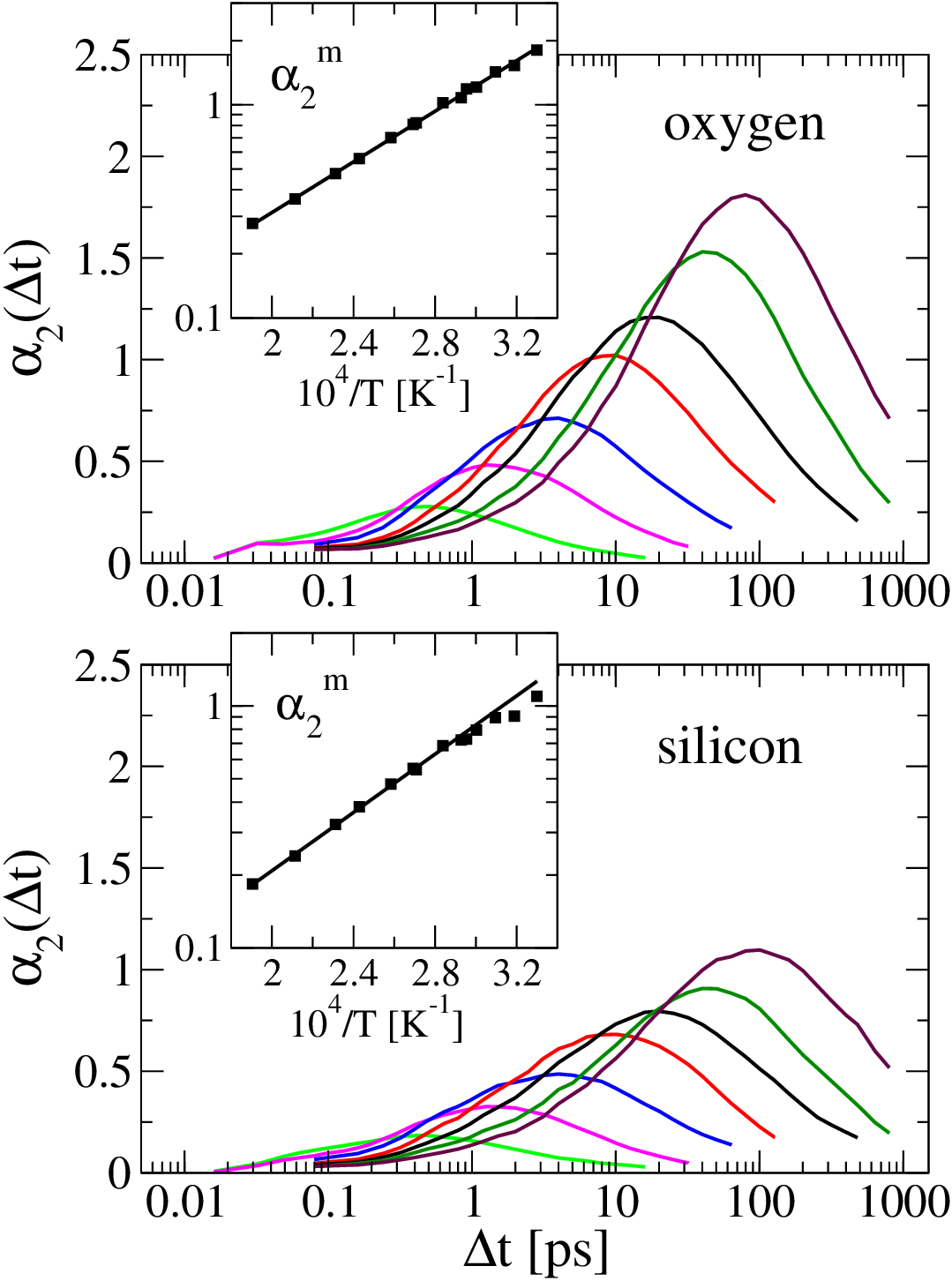}
\caption{Non-Gaussian parameter $\alpha_2(\Delta t)$ for the oxygen
and silicon atoms at various temperatures ($5250\,\mathrm{K}$,
$4330\,\mathrm{K}$, $3870\,\mathrm{K}$, $3520\,\mathrm{K}$,
$3330\,\mathrm{K}$, $3140\,\mathrm{K}$, $3030\,\mathrm{K}$). The
insets show the temperature dependence of the maximum values,
$\alpha_{2}^m$. The solid lines are interpolations of the data at
$T\!>\!3600\,\mathrm K$ with an Arrhenius law. The limited fitting
range was chosen to determine whether $\alpha_{2}^m$ shows a
significantly different temperature dependence at low
temperatures.}\label{A2}
\end{figure}

The deviations from Gaussian behavior can be quantified by the
non-Gaussian parameter \cite{CD-PRE,MA}
\begin{equation}
\alpha_2(\Delta t)=\frac{3}{5} \frac{\langle\,r_i^4(\Delta
 t)\,\rangle}
{\langle\,r_i^2(\Delta t)\,\rangle^2}-1.
\end{equation}
In Fig.~\ref{A2}, we see that $\alpha_2(\Delta t)$ exhibits a
maximum at intermediate times. For both atomic species, the
position of the maximum, $t_{\alpha 2}$, shifts to longer times
and the maximum value, $\alpha_{2}^m\!\equiv\!\alpha_2(t_{\alpha
2})$, increases with decreasing $T$. The insets of Fig.~\ref{A2}
demonstrate that the increase of $\alpha_2^m$ can be described by
an exponential growth with $1/T$. For the silicon atoms, the
temperature dependence may be weaker at $T\!\lesssim\!T_{MCT}$,
but statistics are too poor to draw unambiguous conclusions. In
any case, the data show that the crossover observed for the
transport coefficients is not of great importance for the
temperature dependence of $\alpha_2^m$. At any given $T$, the
non-Gaussian parameter is larger for oxygen than for silicon,
implying that dynamic heterogeneity is more pronounced for oxygen.
The temperature dependence of $t_{\alpha 2}$ is displayed in
Fig.~\ref{TAU}. Upon cooling, $t_{\alpha 2}$ decouples from
$\langle \tau \rangle$ so that the time constants differ by about
one order of magnitude at $T\!=\!3030\,\mathrm K$. Comparison with
Fig.~\ref{FS} shows that the deviations from Gaussian behavior are
largest in the late-$\beta$/ early-$\alpha$ relaxation regime,
consistent with previous results for various models of supercooled
liquids.~\cite{WK,CD-PRE,YG-PO,MA,YG-DZ2,AH-PC,NG}

\subsection{Properties of highly mobile particles}\label{Mobile}

In this section, we investigate the properties of highly mobile
particles on intermediate time scales. Since we are interested in
a detailed comparison with results for non-networked glass forming
liquids, our analysis focuses on quantities studied in previous
work.~\cite{CD-PRE,SCG,YG-DZ2,YG-PO,NG} There, fractions of mobile
particles $\phi\!=\!5\!-\!7\%$ were considered. These fractions
were obtained as the fractions of particles in the tail of
$G_s(r,t_{\alpha 2})$. Note that these highly mobile particles
show displacements of at least one interparticle distance, while
the rest of the particles are trapped in their local cages, see
Fig.~\ref{VAN}. Here, we compare results for $\phi\!=\!3\%$,
$\phi\!=\!5\%$ and $\phi\!=\!7\%$, where the most mobile particles
in a time interval $\Delta t$ are identified based on the scalar
particle displacements within this time window.  We analyze the
behavior of the oxygen and silicon atoms separately, since the
atomic species exhibit somewhat different properties, as shown
below.

\subsubsection{Clusters}

\begin{figure}
\includegraphics[angle=0,width=7.5cm,clip]{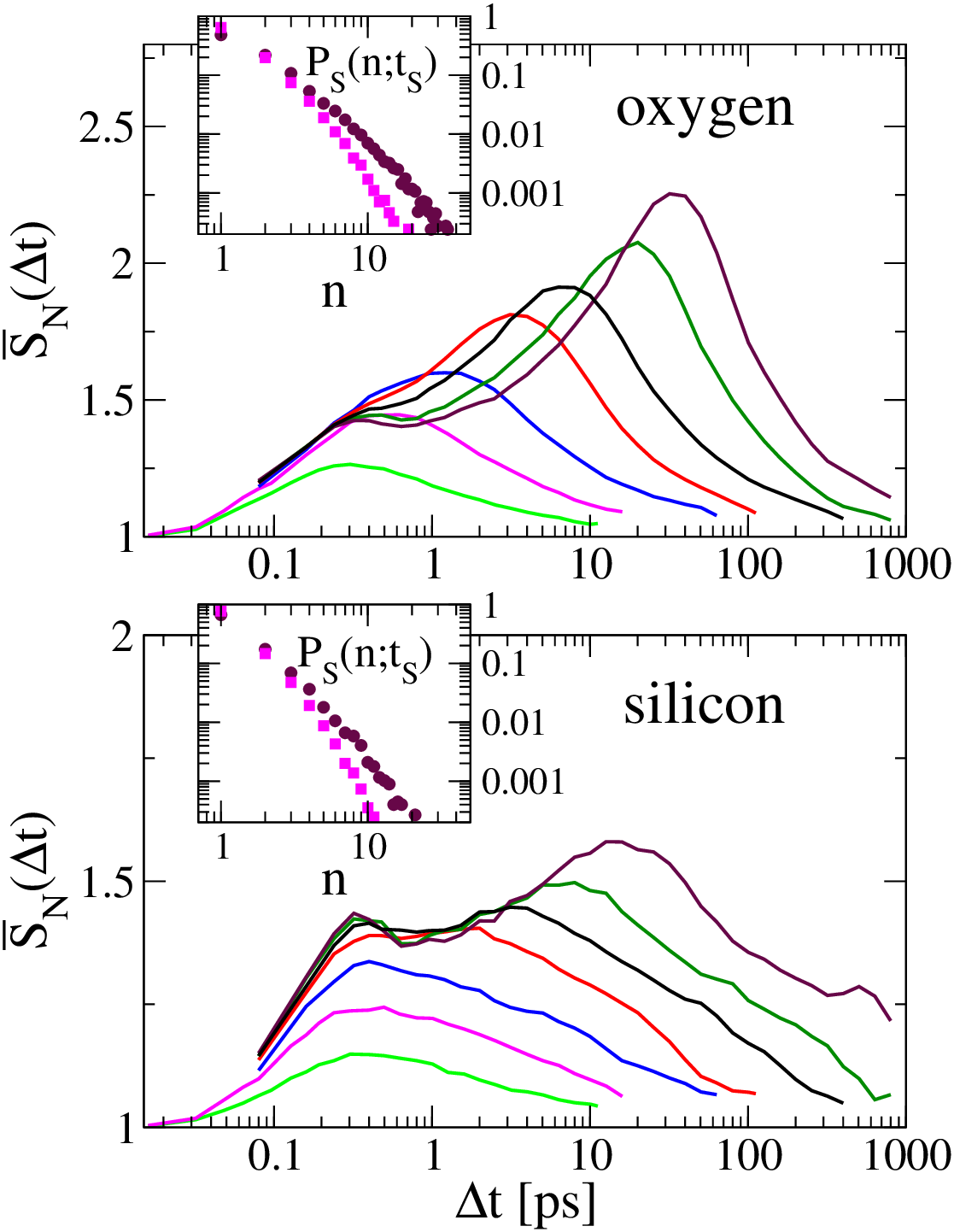}
\caption{Normalized number-averaged mean cluster size,
$\overline{S}_N(\Delta t)$, for the oxygen and silicon atoms at
various temperatures ($5250\,\mathrm{K}$, $4330\,\mathrm{K}$,
$3870\,\mathrm{K}$, $3520\,\mathrm{K}$, $3330\,\mathrm{K}$,
$3140\,\mathrm{K}$, $3030\,\mathrm{K}$) and a fraction
$\phi\!=\!5\%$. The insets show the probability distribution of
the cluster size, $P_S(n;\Delta t)$, for $T\!=\!3030\,\mathrm K$
and $T\!=\!4330\,\mathrm K$, and times when the mean cluster size
is a maximum ($\Delta t\!=\!t_S$).}\label{cluster}
\end{figure}

To ascertain the spatially heterogeneous nature of dynamics in BKS
silica we first show that highly mobile particles form clusters
larger than expected from random statistics. Following previous
studies,~\cite{CD-PRE,SCG,YG-DZ2,YG-PO,NG,MV-PRL} we define a
cluster as a group of highly mobile particles that reside in the
first neighbor shells of each other. For both atomic species, we
define the neighbor shell based on the first minimum of the
respective pair correlation function. A statistical analysis of
the clusters is possible when we determine the probability
distribution $P_S(n;\Delta t)$ of finding a cluster of size $n$
for a time interval $\Delta t$. From this distribution, the
number-averaged and weight-averaged mean cluster size can be
calculated according to
\begin{equation}\label{NA}
S_{N}(\Delta t)=\sum_n\,n\,P_S(n;\Delta t)
\end{equation}
and
\begin{equation}\label{WA}
S_{W}(\Delta t)=\frac{\sum_n\,n^2\,P_S(n;\Delta
t)}{\sum_n\,n\,P_S(n;\Delta t)},
\end{equation}
respectively, where $\sum_n P_S(n)\!=\!1$.

In the following analysis of the clusters and, later, of the
strings, we focus on the number-averaged data for the fraction
$\phi\!=\!5\%$. Here, the number average is considered due to the
smaller statistical error of this quantity. However, we carefully
checked that our conclusions depend neither on the average nor the
fraction considered. For several examples, this is demonstrated by
including results for $\phi\!=\!3\%$ and $\phi\!=\!7\%$. Some
findings for the weight-averaged data were shown in previous
work.~\cite{MV-PRL} They are discussed in section~\ref{disc}, when
we compare SHD in various models of viscous liquids on a
quantitative level. Finally, we determined that an analysis of
clusters consisting of both mobile oxygen and silicon atoms does
not yield new insights.

In Fig.~\ref{cluster}, we display the temperature dependence of
the normalized mean cluster size $\overline{S}_N(\Delta
t)\!=\!S_N(\Delta t)/S_N^*$, where $S_N^*\!=\!1.21$ is the mean
cluster size resulting when 5\% of the particles are randomly
chosen to construct clusters. Thus, $´\overline{S}_N(\Delta t)$
measures exclusively effects due to SHD. Inspecting the oxygen
data, we see that $\overline{S}_N(\Delta t)$ exhibits a peak,
indicating the existence of clusters that are transient in nature.
This peak grows and shifts to longer times upon cooling. In
addition, there is a shoulder near $\Delta
t\!\approx\!0.3\,\mathrm{ps}$. These findings confirm our previous
results for the corresponding weight-averaged data,
$\overline{S}_W(\Delta t)$.~\cite{MV-PRL} For silicon, the
shoulder becomes a separate maximum and $\overline{S}_N(\Delta t)$
shows two peaks at low $T$. Since the position and height of the
primary maximum depend more strongly on temperature than those of
the secondary maximum, both peaks merge at high $T$. Comparing the
data for both atomic species, we observe that the clusters are
larger for oxygen, consistent with a larger value of $\alpha_2^m$
for this species.

A shoulder in $\overline{S}_N(\Delta t)$ at the crossover from the
ballistic to the plateau regime was also observed for
water,~\cite{NG} whereas such an increase is absent in simple
liquids~\cite{CD-PRE,SCG,YG-DZ2} and a polymer melt.~\cite{YG-PO}
In the case of water,~\cite{NG} this phenomenon was ascribed to
``strong correlations in the vibrational motion of first-neighbor
molecules, owing to the presence of hydrogen bonds''. Similarly,
we suggest that the network structure of silica enables a
collective vibrational mode, which becomes more relevant when $T$
is decreased. This conclusion is corroborated by the oscillatory
behavior of $F_s(q,\Delta t)$ in the corresponding time regime,
see Fig.~\ref{FS}. Provided these findings for silica can be
attributed to the Boson peak, as proposed by Horbach \emph{et
al}.,~\cite{JH-BP1,JH-BP2} our findings imply that this phenomenon
results from a local, collective vibration.

\begin{figure}
\includegraphics[angle=0,width=6cm,clip]{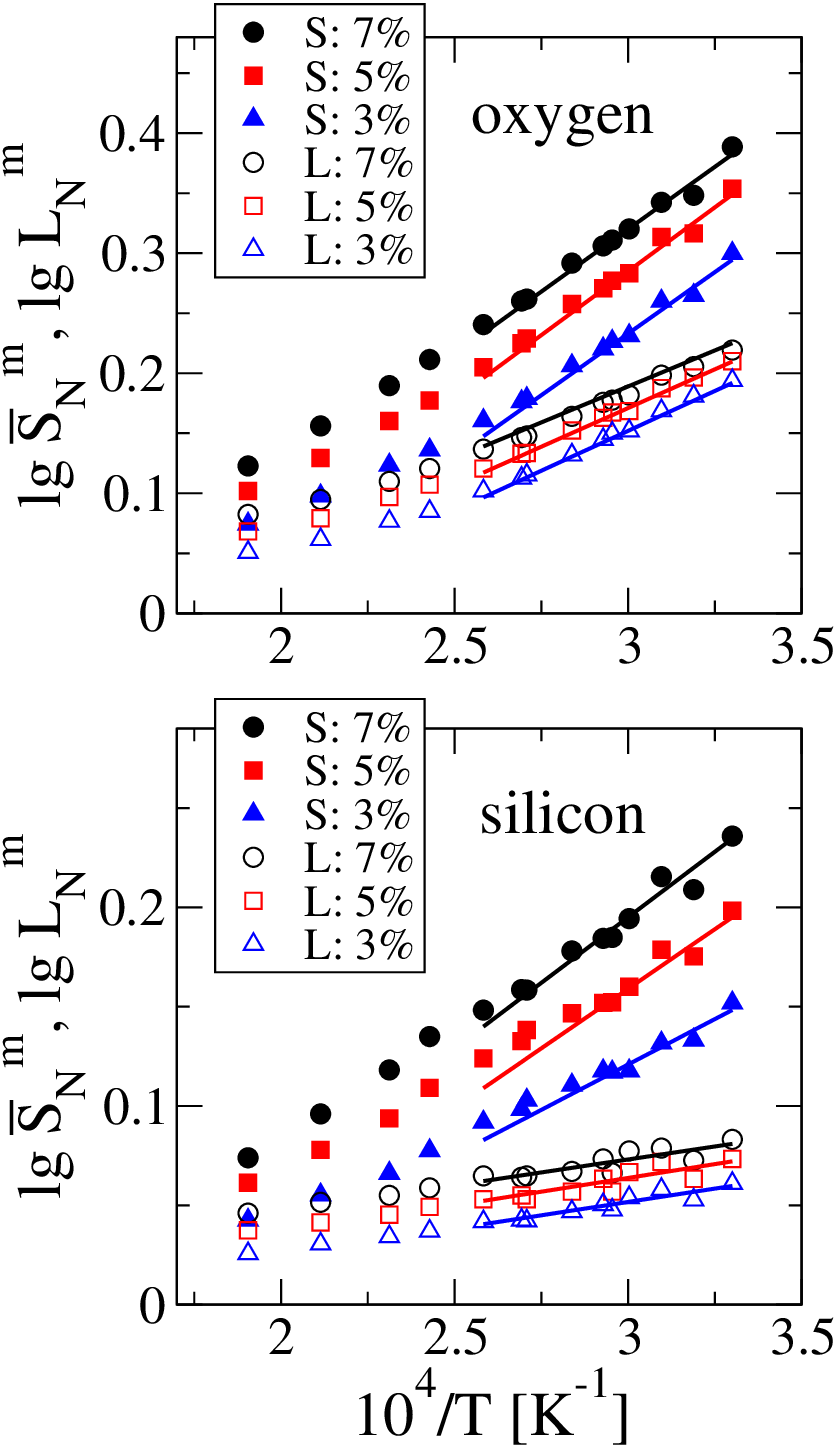}
\caption{Maximum of the normalized mean cluster size,
$\overline{S}_N^{m}$, together with the maximum of the mean string
length, $L_N^m$, for the oxygen and silicon atoms as a function of
the reciprocal temperature. The various fractions of mobile
particles, $\phi$, used in the analysis are indicated. The solid
lines are guides to the eye.}\label{max}
\end{figure}

SHD related to the structural relaxation manifests itself in the
primary maximum of $\overline{S}_N(\Delta t)$. For a quantitative
analysis, we extract the peak times and peak heights. In
Fig.~\ref{TAU}, we see that the mean cluster size is maximum at
times $t_S\!\ll\!\langle \tau \rangle$, where the ratio $\langle
\tau \rangle/t_S$ is nearly independent of $T$. For the silicon
atoms, such analysis is not possible at sufficiently high $T$,
because the primary maximum is submerged by the secondary maximum
so that the temperature-independent position of the latter results
in a constant value for $t_S$. Despite slight deviations, the
temperature dependence of $t_S$ for different values of $\phi$ is
comparable. Specifically, $t_S$ increases with increasing
percentage, but the data for $\phi\!=\!3\%$ and $\phi\!=\!7\%$
differ by less than a factor of 2 at all studied $T$. In
Fig.~\ref{max}, we show the temperature dependence of the peak
height $\overline{S}_N^{m}\!\equiv\!\overline{S}_N(t_S)$ for all
considered fractions $\phi$. Inspecting the data for the lower
$T$, we observe that $\overline{S}_N^{m}$ grows nearly
exponentially with increasing $1/T$. Thus, even in a temperature
range where the relaxation time $\langle \tau \rangle$ varies
according to an Arrhenius law, the cluster size strongly
increases. A more detailed analysis together with a discussion of
the results will be presented in section~\ref{SAG}.

Finally, we investigate the probability distribution of the
cluster size, $P_S(n;\Delta t)$. For simple
liquids~\cite{CD-PRE,SCG,YG-DZ2} and a polymer melt,~\cite{YG-PO}
this distribution exhibits a power-law behavior when $\Delta
t\!=\!t_S$ and $T\!\approx\!T_{MCT}$. In the insets of
Fig.~\ref{cluster}, we show $P_S(n;t_S)$ for the oxygen and
silicon atoms at representative values of $T$. At
$T\!=\!3030\mathrm{\,K}$, clusters containing as many as 40 oxygen
atoms and 20 silicon atoms are observed. However, a power-law is
not found in the studied temperature range, indicating that this
feature of SHD in simple liquids cannot be generalized to the case
of the network-former BKS silica. The data can be satisfactorily
fit by a power-law multiplied by an exponential cutoff. Such
behavior is expected when a percolation transition is approached
and, hence, we cannot exclude that such a transition occurs at
much lower $T$.

\subsubsection{Strings}

It has been demonstrated~\cite{CD-PRL,CD-PRE,YG-DZ2,MA} that
dynamics in several fragile liquids are facilitated by string-like
motion, i.e., groups of particles follow each other along
one-dimensional paths. However, our previous work~\cite{MV-PRL}
suggests that this type of motion is less important for silica. To
study the relevance of string-like motion in more detail, we
follow Donati \emph{et al.\ }\cite{CD-PRL} and construct strings
by connecting any two particles $i$ and $j$ of the same atomic
species if
\begin{displaymath}
\mathrm{min}[\,|\vec{r}_{i}(t_0)\!-\!\vec{r}_{j}(t_0\!+\!\Delta
t)|,|\vec{r}_{i}(t_0\!+\!\Delta t)\!-\!\vec{r}_{j}(t_0)|\,]
\!<\!\delta.
\end{displaymath}
Similar to the values used in previous
work~\cite{CD-PRL,CD-PRE,YG-DZ2,MA}, we set $\delta$ to $\sim 55\%$
of the respective interatomic distance, resulting in
$\delta\!=\!1.4\mathrm{\,\AA}$ and $\delta\!=\!1.7\mathrm{\,\AA}$ for
oxygen and silicon, respectively. Then, the above condition implies
that one particle has moved and another particle has occupied its
position. We checked that our conclusions are not affected when
$\delta$ is varied in a meaningful range.

\begin{figure}
\includegraphics[angle=0,width=7.5cm,clip]{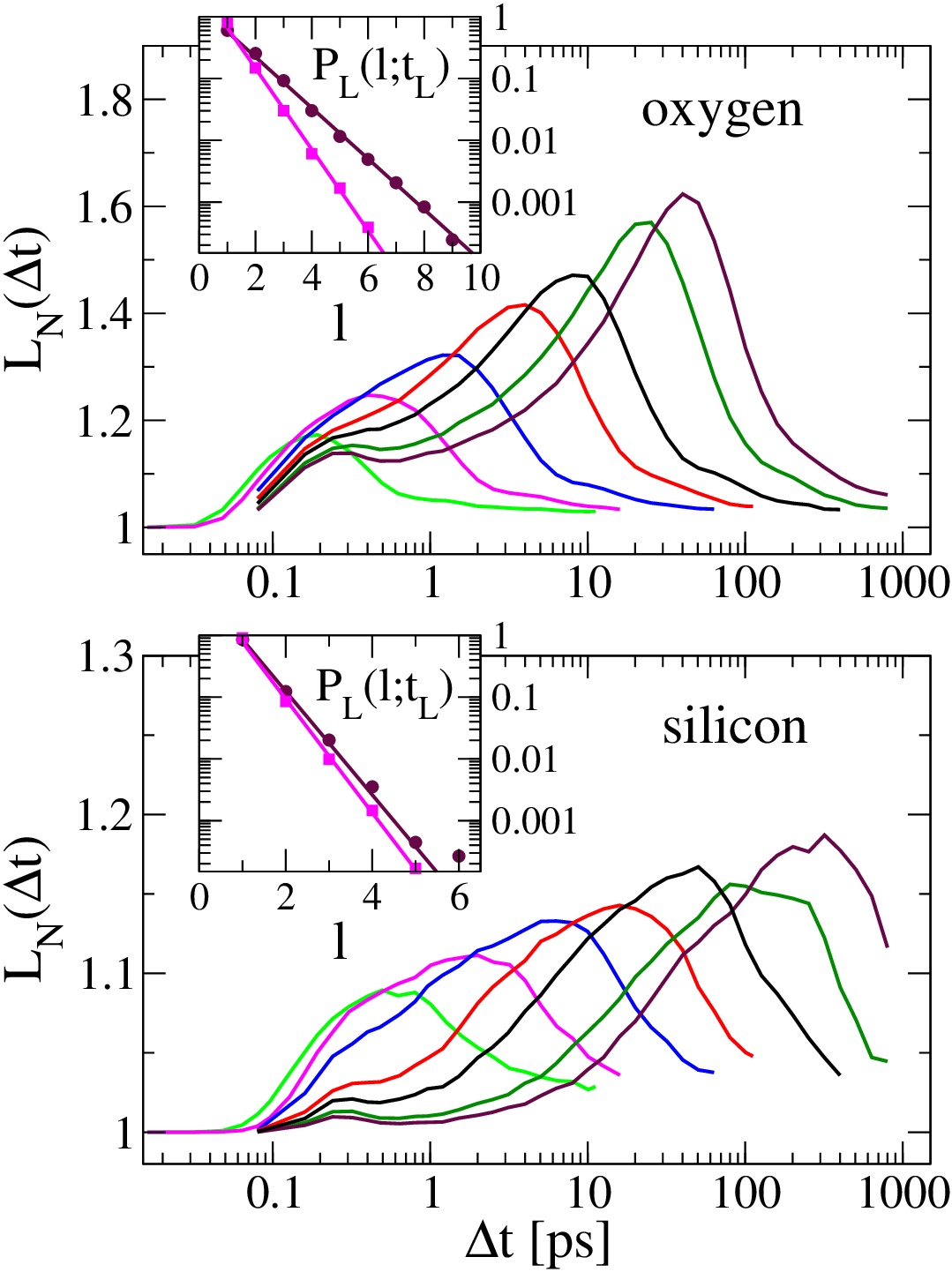}
\caption{Number-averaged mean string length, $L_N(\Delta t)$, for
the oxygen and silicon atoms at various temperatures
($5250\,\mathrm{K}$, $4330\,\mathrm{K}$, $3870\,\mathrm{K}$,
$3520\,\mathrm{K}$, $3330\,\mathrm{K}$, $3140\,\mathrm{K}$,
$3030\,\mathrm{K}$), where $\phi\!=\!5\%$. The insets show the
probability distribution of the string length, $P(l;\Delta t)$,
for $T\!=\!3030\,\mathrm K$ and $T\!=\!4330\,\mathrm K$, and times
when the mean string size is a maximum ($\Delta
t\!=\!t_L$).}\label{string}
\end{figure}

For an analysis of string-like motion, we determine the
probability distribution $P_L(l;\Delta t)$ of finding a string of
length $l$ in a time interval $\Delta t$ and calculate the
number-averaged mean string length $L_N(\Delta t)$ in analogy with
Eq.~\ref{NA}. Note that we include ``strings'' containing only one
atom. In Fig.~\ref{string}, we display the temperature dependence
of $L_N(\Delta t)$. Though the strings for both atomic species
grow and shrink in time, string-like motion is very different for
the oxygen and silicon atoms. We see that, at any given $T$,
$L_N(\Delta t)$ is maximum at a much later time for silicon than
for oxygen. Further, the mean string length is significantly
smaller for the former than for the latter atomic species. In
particular, for the silicon atoms, the small values $L_N(\Delta
t)\!\approx\!1$ imply very limited string-like motion.

The different behavior of the oxygen and silicon atoms can be
quantified by the peak times $t_L$ and peak heights
$L_N^m\!\equiv\!L_N(t_L)$. The former are included in
Fig.~\ref{TAU}. For oxygen, we find that the mean string length
and the mean cluster size are maximum at similar times
$t_S\!\approx\!t_L\!\ll\!\langle \tau \rangle$, as was observed
for simple liquids~\cite{CD-PRE,YG-DZ2} and a polymer
melt.~\cite{MA} In contrast, these mean sizes peak at very
different times for silicon where the few short strings are
largest in the $\alpha$-relaxation regime, i.e.,
$t_S\!\ll\!t_L\!\approx\!\langle \tau \rangle$. This finding
clearly shows that the string-like motion known for fragile
liquids is suppressed for the silicon atoms. In Fig.~\ref{max}, we
show the temperature dependence of the peak heights $L_N^m$ for
all studied fractions $\phi$. As was observed for the clusters,
the mean string size for the oxygen atoms increases nearly
exponentially with $1/T$ at the lower $T$. For the silicon atoms,
the strings grow more slowly, but large scattering of the data, in
particular at low $T$, hampers a quantification of the temperature
dependence.

Examples of the probability distribution of the string length,
$P_L(l,\Delta t)$, are shown in the insets of Fig.~\ref{string}.
For all considered $T$ and both atomic species, $P_L(l,\Delta t)$
decays exponentially when $\Delta t\!=\!t_L$, as was observed in
previous studies of string-like
motion.~\cite{CD-PRL,CD-PRE,YG-DZ2,MA} This finding suggests that
an exponential distribution $P_L(n,t_L)$ is a feature inherent to
strings constructed in the way described above. In view of the
exponential distribution, an analogy to the equilibrium
polymerization of linear polymers was proposed in previous
work.~\cite{CD-PRL,MA}

\begin{figure}
\includegraphics[angle=0,width=7.5cm,clip]{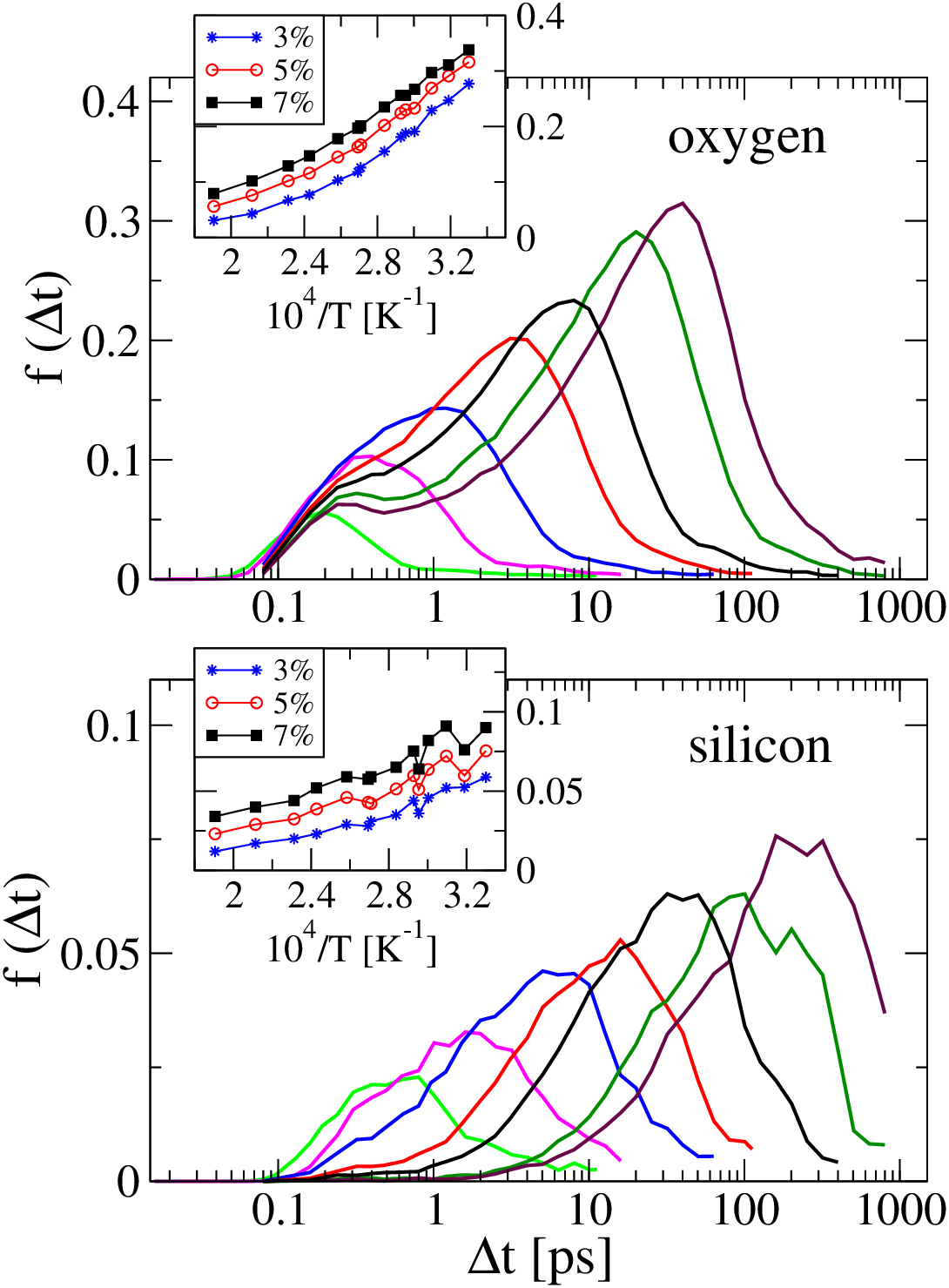}
\caption{Fraction $f(\Delta t)$ of highly mobile oxygen and silicon
atoms, respectively, that are involved in non-trivial string-like
motion ($l\!\geq\!3$). We display results for $5250\,\mathrm{K}$,
$4330\,\mathrm{K}$, $3870\,\mathrm{K}$, $3520\,\mathrm{K}$,
$3330\,\mathrm{K}$, $3140\,\mathrm{K}$ and  $3030\,\mathrm{K}$. The
insets show the maximum value, $f^m$, as a function of the reciprocal
temperature. The values of $\phi$ used in the analysis are
indicated.}\label{ratio}
\end{figure}

Finally, we analyze the relevance of string-like motion for the
relaxation of the most mobile particles in silica. Following previous
work,~\cite{YG-DZ2} we quantify the importance by the fraction of
mobile particles, $f(\Delta t)$, that are involved in non-trivial
strings, i.e., strings with $l\!\geq\!3$. In Fig.~\ref{ratio}, we see
that string-like motion is of very limited relevance for silicon,
whereas it becomes increasingly important for oxygen upon cooling.
Hence, this dynamical pattern may be an important channel of
relaxation for oxygen near $T_g$. A more detailed analysis reveals
that $f(\Delta t)$ is maximum at times $\Delta
t\!=\!t_f\!\approx\!t_L$ for both the oxygen and the silicon atoms,
see Fig.~\ref{TAU}. Inspecting the maximum amplitude $f^m$ in the
insets of Fig.~\ref{ratio}, it is evident that our conclusions do not
depend on the value of $\phi$.

\subsection{Relation to the Adam-Gibbs theory}\label{SAG}

\begin{figure}
\includegraphics[angle=0,width=7.5cm,clip]{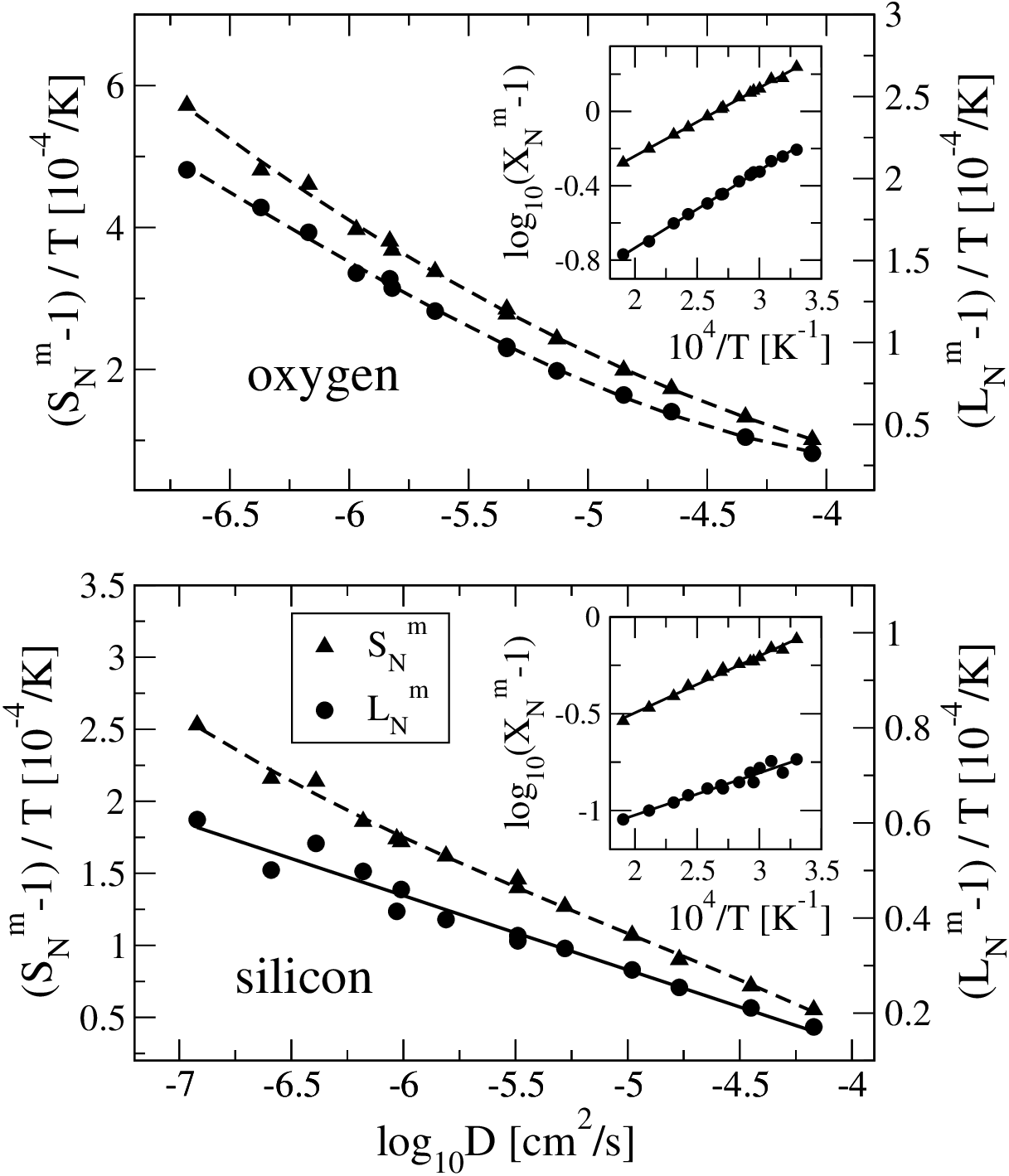}
\caption{$S_N^{m}\!-\!1$ and $L_N^{m}\!-\!1$ as a function of the
logarithm of the diffusion coefficient $D$. The dashed lines are
guides to the eye. The solid line is a linear interpolation, which
will be expected if clusters or strings are related to the
cooperatively rearranging regions of the Adam-Gibbs
theory.~\cite{NG,YG-PHD} In the insets, we depict the temperature
dependence of
$X_N\!-\!1\!\equiv\!\overline{S}_N^{m}\!-\!1,L_N^{m}\!-\!1$
together with Arrhenius fits (solid lines).}\label{ag}
\end{figure}

The Adam-Gibbs (AG) relation~\cite{AG} $D\!\propto\!\exp[-A/(TS_c)]$,
which links the diffusion coefficient with the configurational
entropy $S_c$, was found to hold for a BLJ liquid,~\cite{SS}
water~\cite{AS} and the present case of BKS silica.~\cite{SV} The AG
theory further proposes a relation between $S_c$ and the
characteristic mass of cooperatively rearranging regions (CRR).
However, the CRR are not precisely defined in the theory and their
nature is still elusive. In recent work on water, Giovambattista
\emph{et al.}~\cite{NG} observed that
\begin{equation}\label{CSC}
(S_N^{m}\!-\!1)\propto1/S_c,
\end{equation}
suggesting that, first, the (non-normalized) mean cluster size is a
measure of the mass of the CRR and, second, a cluster of size one
does not correspond to a CRR. Furthermore, they confirmed along the
considered isochore the expectation
\begin{equation}\label{AGC}
D\propto\exp[-A^*(S_N^m-1)/T]
\end{equation}
resulting from the AG relation together with Eq.~\ref{CSC}.

Here, we analyze whether Eq.~\ref{AGC} is valid for BKS silica.
Since the AG relation holds for this model in the studied
temperature range,~\cite{SV} the validity (failure) of
Eq.~\ref{AGC} implies the validity (failure) of Eq.~\ref{CSC} and,
hence, we can determine whether a relation between the clusters of
mobile particles and the CRR exists, as was reported for
water.~\cite{NG} To this end, we extract the diffusion coefficient
from the long-time behavior of the mean square
displacement~\cite{MV-PRL} and plot $(S_N^m\!-\!1)/T$ as a
function of $\log D$ in Fig.~\ref{ag}. Evidently, a linear
relation is observed for neither oxygen nor silicon, indicating a
failure of Eq.~\ref{AGC} and, thus, also Eq.~\ref{CSC}. Hence, for
silica, the mean cluster size $S_N^m$ is not a measure of the
characteristic mass of the CRR. We also find that such relation is
not valid for the weight-averaged data $S_W^m$ or for clusters
consisting of both oxygen and silicon atoms. On the other hand, it
has been argued~\cite{SCG,YG-DZ2,YG-PHD} that the strings and not
the clusters are the elementary units of cooperative motion and,
hence, related to the CRR. Therefore, we include $(L_N^m\!-\!1)/T$
as a function of $\log D$ in Fig.~\ref{ag}. While no linear
relation is found for the oxygen atoms, such behavior cannot be
ruled out for the silicon atoms. However, the variation of $L_N^m$
is too small to draw any conclusion in the latter case, where
string-like motion is peculiar and largely insignificant anyway.

In the insets of Fig.~\ref{ag}, we show the temperature dependence
of $S_N^m\!-\!1$ and $L_N^m\!-\!1$. For both atomic species, we
observe that an exponential growth with $1/T$ fits the data well
over the entire temperature range. This behavior is confirmed by
our analysis for the fractions $\phi\!=\!3\%$ and $\phi\!=\!7\%$
(not shown), indicating that, in a strong glass former, SHD
becomes increasingly important with decreasing $T$. In particular,
we find no evidence for substantial effects due to the
``fragile-to-strong crossover'' observed for the transport
coefficients. Provided the exponential growth continues upon
further cooling, the mean cluster and mean string size are finite
at all finite $T$. In contrast, a power-law temperature dependence
of the mean cluster size was found for a BLJ
liquid,~\cite{CD-PRE,SCG} suggesting a percolation transition of
clusters of mobile particles in the vicinity of $T_{MCT}$.

\subsection{Behavior of four-point spatiotemporal density fluctuations}

\begin{figure}
\includegraphics[angle=0,width=6cm,clip]{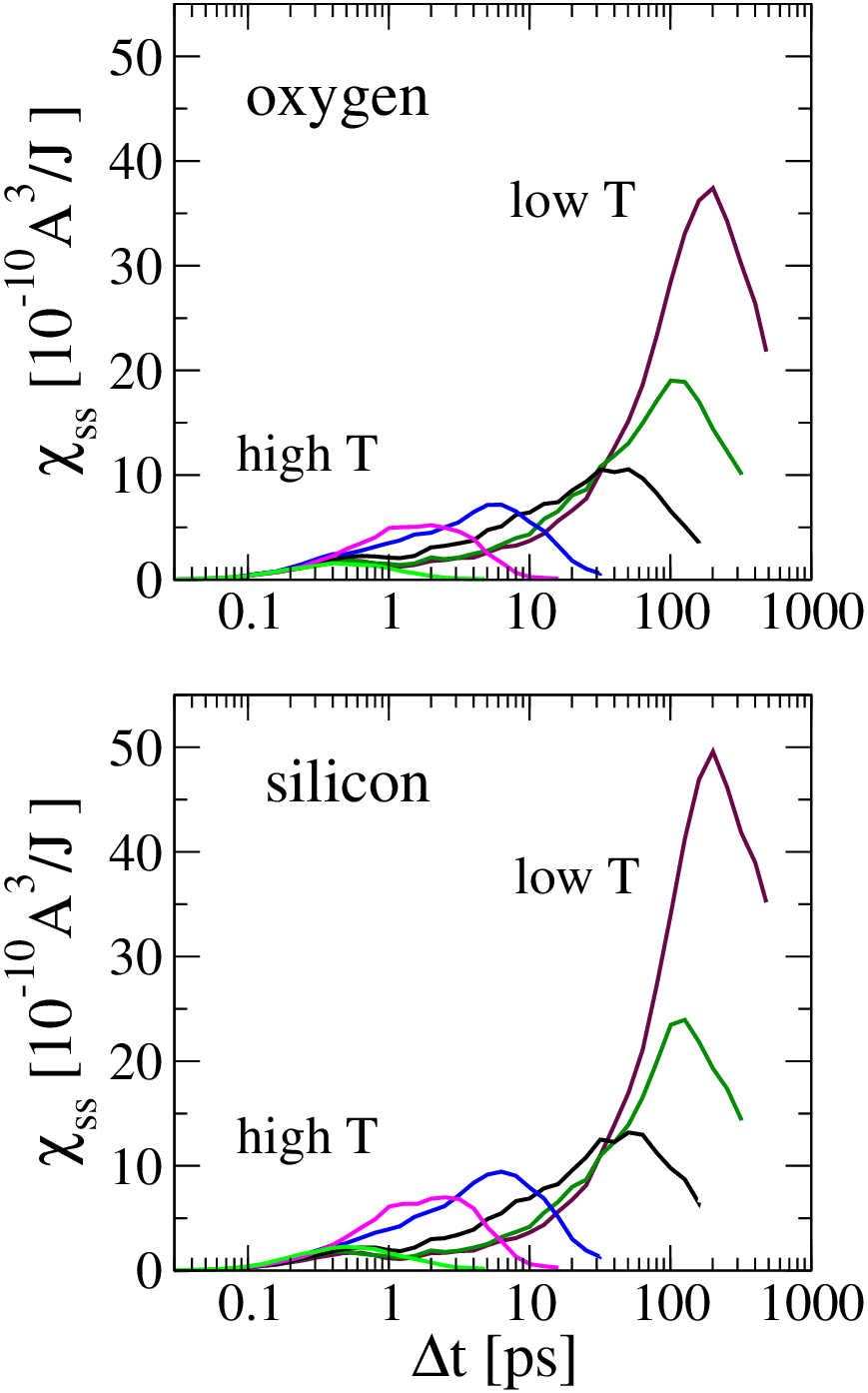}
\caption{Generalized susceptibility $\chi_{ss}(\Delta t)$ for the
oxygen and silicon atoms. The temperatures are $5250\,\mathrm{K}$,
$4330\,\mathrm{K}$, $3870\,\mathrm{K}$, $3330\,\mathrm{K}$,
$3140\,\mathrm{K}$ and $3030\,\mathrm{K}$.}\label{chi}
\end{figure}

Previous work~\cite{SCG-JCP,NL,NL2,LB,CDA} showed that two-point,
two-time fourth order density correlation functions are well
suited to study SHD. The general theoretical framework is
described in detail in the literature.~\cite{NL2,SF} In brief, it
has been proven useful to define an ``order parameter'' $Q(\Delta
t)$ that compares the configurations of a liquid with density
$\rho(\vec{r},t)\!=\!\sum_{i=1}^N\delta(\vec{r}\!-\!\vec{r}_i(t))$
at two different times $\Delta t\!\equiv\!t_2\!-\!t_1$
\begin{eqnarray}
Q(\Delta t)&=&\int d^3r_1 d^3r_2 \;\rho(\vec{r}_1,0)
\rho(\vec{r}_2,\Delta t)\cos\,[\,\vec{q}(\vec{r}_1\!-\!\vec{r}_2)]\nonumber\\
&=&\sum_{i=1}^N\sum_{j=1}^N\cos\,[\,\vec{q}(\vec{r}_i(0)\!-\!\vec{r}_j(\Delta
t))].
\end{eqnarray}
$Q(\Delta t)$ counts the number of ``overlapping'' particles in
two configurations separated by a time interval $\Delta t$. Here,
we do not apply a strict cutoff to define overlapping particles,
as was done in previous studies,~\cite{SCG-JCP,NL,NL2} but we
follow Berthier~\cite{LB} and use the intermediate scattering
function, $\cos\,[\,\vec{q}(\vec{r}_1\!-\!\vec{r}_2)]$, where we
again choose $q\!=\!1.7\mathrm{\,\AA^{-1}}$. The fluctuations of
this quantity are described by a generalized susceptibility
\begin{equation}
\chi_4(\Delta t)=\frac{\beta V}{N^2}\left[\langle Q^2(\Delta
t)\rangle-\langle Q(\Delta t)\rangle^2\right],
\end{equation}
which corresponds to the volume integral of a four-point density
correlation function
\begin{eqnarray}
\chi_4(\Delta t)&=& \frac{\beta V}{N^2}\int d^3r_1\, d^3r_2\, d^3r_3\, d^3r_4\; \cos\,[\,\vec{q}(\vec{r}_1\!-\!\vec{r}_2)]\nonumber\\
&&\times\cos\,[\,\vec{q}(\vec{r}_3\!-\!\vec{r}_4)]\;G_4(\vec{r}_1,\vec{r}_2,\vec{r}_3,\vec{r}_4,\Delta
t)
\end{eqnarray}
where $\beta\!=\!k_BT$ and
\begin{eqnarray}
G_4(\vec{r}_1,\vec{r}_2,\vec{r}_3,\vec{r}_4,
t)&=&\langle\rho(\vec{r}_1,0)\rho(\vec{r}_2, t)\rho(\vec{r}_3,0)\rho(\vec{r}_4, t)\rangle\nonumber\\
&&-\langle\rho(\vec{r}_1,0)\rho(\vec{r}_2,
t)\rangle\langle\rho(\vec{r}_3,0)\rho(\vec{r}_4, t)\rangle.\nonumber
\end{eqnarray}
The overlapping particles are comprised of localized particles
that have hardly moved and particles that have been replaced by a
neighboring particle. Accordingly, $\chi_4$ can be decomposed into
self- ($\chi_{SS}$), distinct- ($\chi_{DD}$) and interference
($\chi_{SD}$) parts, where the main contribution was found to
result from the self part.\cite{SCG-JCP,NL2} In this section, we
are interested in spatial correlations of highly immobile
particles. Therefore, we focus on the self part, which is given by
\begin{equation}
\chi_{SS}(\Delta t)=\frac{\beta V}{N^2}\left[\langle Q_S^2(\Delta
t)\rangle-\langle Q_S(\Delta t)\rangle^2\right],
\end{equation}
where
\begin{equation}
Q_S(\Delta t)=\sum_{i=1}^N
\cos\,[\,\vec{q}(\vec{r}_i(0)\!-\!\vec{r}_i(\Delta t))].
\end{equation}

In Fig.~\ref{chi}, we show $\chi_{SS}(\Delta t)$ for the oxygen
and silicon atoms at several values of $T$. While the
susceptibility is small at high $T$, it shows a pronounced maximum
at low $T$. This strong increase of the peak height, $\chi_4^m$,
indicates a growing range of spatial correlations between immobile
particles with decreasing $T$.~\cite{SCG-JCP,NL,NL2,LB}
Unfortunately, a detailed analysis of the temperature dependence
of $\chi_4^m$ is hampered by a large scattering of the values (not
shown). To characterize the times when the fluctuations in the
number of localized particles are biggest, we extract the peak
times $t_4$. From the results in Fig.~\ref{TAU}, it is evident
that, for both atomic species, the spatial correlations between
highly immobile particles are maximum in the $\alpha$-relaxation
regime. In particular, $\langle\tau\rangle$ and $t_4$ show a very
similar temperature dependence, where the ratio
$\langle\tau\rangle/t_4$ amounts to a factor of about two. On the
other hand, $t_4$ is about one order of magnitude longer than
$t_S$ and, hence, SHD analyzed from the perspectives of highly
immobile and highly mobile particles, respectively, is most
pronounced at different times. All these findings are consistent
with previous results for BLJ liquids.~\cite{SCG-JCP,NL,NL2,LB,SF}

\subsection{Relation between structure and dynamics}

\begin{figure}
\includegraphics[angle=0,width=7.5cm,clip]{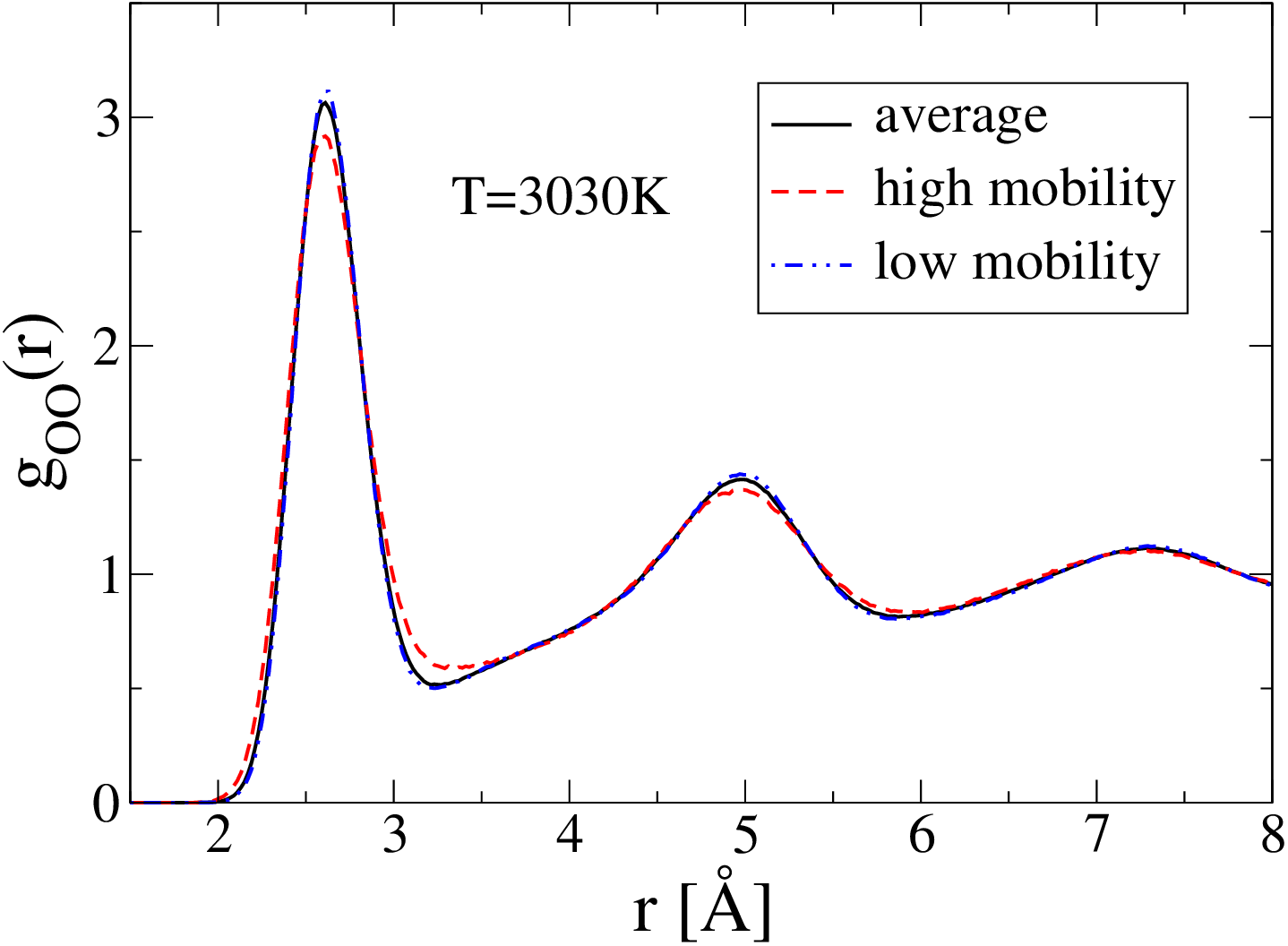}
\caption{Pair correlation functions $g_{OO}(r)$ at
$T\!=\!3030\mathrm{\,K}$. We compare results obtained for all oxygen
atoms with those for the most mobile and the most immobile oxygen
atoms in a time interval $\Delta
t\!=\!10\mathrm{\,ps}\!\approx\!t_S$, respectively.}\label{g(r)}
\end{figure}

Amorphous silica consists of a network of well-defined silicate
tetrahedra. Hence, one may speculate that high particle mobility
is found where the local structure is distorted. To study the
relation between structure and dynamics, we identify the most
mobile and the most immobile particles during a time interval
$\Delta t$ and characterize their respective environments at the
beginning of this period by means of the pair correlation
functions (PCF), $g_{\alpha\beta}(r)$, and the distributions of
coordination numbers, $z_{\alpha\beta}(n)$
($\alpha,\beta\!\in\!\{\mathrm{O,Si}\}$). In Fig.~\ref{g(r)}, we
compare $g_{OO}(r)$ for the mobile and immobile oxygen atoms,
where $\Delta t\!\approx\!t_S$ and $T\!=\!3030\mathrm{\,K}$. While
the data for the immobile particles are comparable to those for
the ensemble average, the first and second neighbor peak are
slightly broadened for the mobile oxygen atoms. This behavior is
similar for all $g_{\alpha\beta}(r)$. Thus, mobile particles
during a time interval $\Delta t_S$ exhibit a somewhat less
ordered local environment at the beginning of this time window, at
least as measured by $g_{\alpha\beta}(r)$, but the effects are
weak. A variation of $\Delta t$ in a meaningful range does not
change these conclusions.

Some of the results from our analysis of the distributions
$z_{\alpha\beta}(n)$ are compiled in Table~\ref{T1}, where the
coordination number $n$ is defined as the number of $\beta$ atoms
for which the distance from a given $\alpha$ atom is smaller than
that corresponding to the first minimum of $g_{\alpha\beta}(r)$.
It is evident that mobile oxygen and silicon atoms exhibit
somewhat higher mean coordination numbers $\overline{z}_{OO}$ and
$\overline{z}_{SiSi}$, respectively, suggesting that their
positions are less favorable in terms of the potential energy.
Further, the percentage of mobile oxygen atoms in the ``ideal''
coordination with two silicon neighbors is smaller than for an
average oxygen atom. Likewise, compared to an average particle,
mobile silicon atoms are more often three- or fivefold coordinated
to oxygen. However, the vast majority of the mobile particles are
still in their preferred local structure at the beginning of the
particular time interval and, hence, defects of the local
structure are not a necessary precondition of high particle
mobility. On the other hand, the observed correlations imply that
a somewhat less ordered local environment facilitates particle
dynamics. Thus, the relation between structure and dynamics in BKS
silica is non-trivial, as was also observed for BLJ
mixtures~\cite{CD-PRE} and the Dzugutov liquid.~\cite{MD}

\begin{table}
\begin{math}
\begin{array}{|c|c|c|} \hline
& \mathrm {average} &  \mathrm{mobile} \\
 \hline
\overline{z}_{OO} & 7.2 & 7.5 \\
\hline
\overline{z}_{SiSi} & 4.2 & 4.4  \\
\hline
z_{OSi}(n\!=\!2) & 0.99 & 0.94 \\
\hline
z_{SiO}(n\!=\!4) & 0.98 & 0.94 \\
\hline
\end{array}
\end{math}
\caption{Coordination numbers characterizing the local structure in
BKS silica at $T\!=\!3030\mathrm{\,K}$.}\label{T1}
\end{table}

\section{Discussion}\label{disc}

\begin{table*}
\begin{math}
\begin{array}{|c|c|c|c|c|c|c|c|} \hline
& \mathrm{BLJ_1} & \mathrm{BLJ_2} &  \mathrm{PM} & \mathrm {DZ} &  \mathrm{H_2O} &  \mathrm{O} &  \mathrm{Si} \\
\hline
\beta & 0.75 & 0.55  & 0.70 & 0.50 & 0.75 & 0.83 & 0.85 \\
\hline
\alpha_2^m & 1.6 & 2.0 & 1.3 & 2.5 & 1.3 & 1.3 & 0.8 \\
\hline
S_W^m & 15\,(5) & & 16\,(6.5) & 22\,(5) & & 7.9\,(7) & 3.9\,(7) \\
\hline
\overline{S}_W^m & & & 6.0\,(6.5)& 11(5) & 4.2\,(7) & 4.9\,(7) & 2.8\,(7) \\
\hline
L_N^m & 2.2\,(5) & & 1.9\,(6.5) & 2.4\,(5) & & 1.5\,(5) & 1.1\,(5) \\
\hline
f^m & & & & 0.70\,(5) & & 0.24\,(5) & 0.07\,(5) \\
\hline
\chi_4^m & 4.7 & 18 & & & & & \\
 \hline
\end{array}
\end{math}
\caption{Quantities characterizing dynamic heterogeneity in
various model liquids at $T\!\approx\!T_{MCT}$. We compare data
for a 80:20 binary Lennard-Jones mixture ($\mathrm{BLJ_1}$,
$T\!=\!1.04\,T_{MCT}$),~\cite{CD-PRL,CD-PRE}, a 50:50 binary
Lennard-Jones mixture ($\mathrm{BLJ_2}$, $T\!=\!T_{MCT}$),
~\cite{TS,NL2} a polymer melt (PM,
$T\!=\!1.02\,T_{MCT}$),~\cite{YG-PO,MA,CB-PRE} the Dzugutov liquid
(DZ, $T\!=\!1.05\,T_{MCT}$)~\cite{YG-DZ2,YG-PHD} and water
($\mathrm{H_2O}$, $T\!=\!1.04\,T_{MCT}$)~\cite{NG,FWS} with the
present results for the oxygen (O) and silicon (Si) atoms in BKS
silica at $T\!=\!T_{MCT}$. The values in brackets denote the
fractions of mobile particles, $\phi$, used in the respective
analysis (in percent).}\label{T2}
\end{table*}

One goal of the present work is to compare dynamic heterogeneity
in various model liquids. For this purpose, we show our results
together with the corresponding literature data in Table~\ref{T2}.
To enable a quantitative comparison, we consider results for
$T\!\approx\!T_{MCT}$ in each case. Inspecting the data, it
becomes clear that there is a correlation between the properties
of dynamic heterogeneity at intermediate times and the features of
the $\alpha$ relaxation, e.g., the stretching parameter $\beta$
characterizing the nonexponentiality of the decay of the
incoherent intermediate scattering function. Specifically, the
Dzugutov liquid shows the most pronounced dynamic heterogeneity at
intermediate times and the smallest value of $\beta$, whereas
systems like silica, with less heterogeneous dynamics at
intermediate times, exhibit a larger $\beta$. In what follows, we
discuss the findings for various model liquids in more detail.

First, we see from Table~\ref{T2} that the non-Gaussian parameter
$\alpha_2$ attains the largest values for the non-network forming
liquids. Though $\alpha_2$ was found to characterize aspects of
the heterogeneity of dynamics,~\cite{AH-PC} it does not provide
direct information about the spatial arrangement of mobile and
immobile particles. The spatially heterogeneous nature of dynamics
in the vicinity of $T_{MCT}$ can be studied by identifying
clusters of mobile particles.~\cite{CD-PRE,YG-PO,YG-DZ2,NG} On a
qualitative level, the clusters in various model liquids,
including BKS silica, show an analogous behavior. For example, the
clusters strongly grow upon cooling, where, at each $T$, the mean
cluster size is maximum at times $t_S$ in the late-$\beta$/
early-$\alpha$ relaxation regime. On a quantitative level, several
findings for simple liquids~\cite{CD-PRE,YG-DZ2} and a polymer
melt,~\cite{YG-PO,MA} e.g., a power-law distribution of the
cluster size, cannot be generalized to the case of silica.
Moreover, there is a spectrum of maximum mean cluster sizes, the
higher and lower end of which are attained by the Dzugutov liquid
and the network formers silica and water, respectively, see
Tab.~\ref{T2}. However, one has to consider that, with the same
definition, see section~\ref{Mobile}, the sizes of the clusters
depend on the properties of the local structure. Specifically,
compared to simple liquids, the particles in network-forming
liquids exhibit smaller coordination numbers and, hence, the
probability that a mobile particle has a mobile neighbor is
smaller. This means that, within the same definition, the clusters
are less likely to grow. At least to some extent, these effects
are canceled out when the mean cluster size is normalized by the
value expected from random statistics. However, at the present
time, it is not clear whether $\overline{S}_{N}$ and
$\overline{S}_{W}$ allow one to study SHD completely independent
of the local structure. In any case, a comparison of the results
among the simple liquids, where the coordination numbers are
similar, suggests a relation between the mean cluster size and the
stretching parameter $\beta$.

A main difference between various model liquids is the relevance
of string-like motion at $T\!\approx\!T_{MCT}$. From the mean
string length $L_N^m$ and the fraction $f^m$ of mobile particles
moving in non-trivial strings, it is evident that this type of
motion is an important channel of relaxation in the Dzugutov
liquid, but not in BKS silica, see Tab.~\ref{T2}. Moreover, the
two atomic species in silica behave differently. On the one hand,
some string-like motion is observed for the oxygen atoms. In
agreement with previous results,~\cite{CD-PRE,MA,YG-DZ2} the mean
string length is maximum at intermediate times
$t_L\!\approx\!t_S$, where the peak height $L_N^m$ increases with
decreasing $T$. Hence, despite a limited relevance at
$T\!\approx\!T_{MCT}$, string-like motion may become important for
the structural relaxation of the oxygen atoms near $T_g$, cf.\
Fig.~\ref{string}. On the other hand, the silicon atoms do
\emph{not} show string-like motion at intermediate times. Instead,
there are very few short strings at much later times
$t_S\!\ll\!t_L\!\approx\!\langle \tau \rangle$, implying that
their nature is different from that observed for other model
liquids.~\cite{CD-PRE,MA,YG-DZ2} Therefore, we conclude that the
``usual'' string-like motion is absent for the silicon atoms. When
we also consider that the oxygen and silicon atoms typically
exhibit two and four covalent bonds, respectively, our findings
imply that string-like motion is suppressed by the presence of
covalent bonds.

Giovambattista \emph{et al.}~\cite{NG} found for water that the
diffusion coefficient $D$ is related to the mean cluster size via the
AG relation where the cluster size is a measure of the mass of the
CRR. In particular, they showed that a linear relation exists between
$\log D$ and $S_N^m\!-\!1$. For silica in the studied temperature
range, we observe no such relation between $D$ and either the mean
cluster size or the mean string size and, hence, the findings for
water cannot be generalized to the case of BKS silica.

An analysis of the generalized susceptibility $\chi_{ss}(\Delta
t)$ revealed spatial correlations of localized, i.e. highly
immobile, particles in BKS silica. On a semi-quantitative level,
our results are in good agreement with that for BLJ
liquids.~\cite{SCG-JCP,NL,NL2,LB} More precisely,
$\chi_{ss}(\Delta t)$ shows a maximum that strongly increases with
decreasing $T$, indicating a growing length of spatial
correlations between highly immobile particles. Moreover, the peak
time $t_4$ is located in the $\alpha$-relaxation regime and
closely follows the temperature dependence of the structural
relaxation. A quantitative comparison of the present values of
$\chi_4^m$ with literature data is hampered by the different units
(reduced/real) used in the simulations. However, the data for the
80:20 BLJ and 50:50 BLJ liquids again imply that the structural
relaxation is more stretched for systems with pronounced SHD.

Very recently, Garrahan and Chandler~\cite{GC-PRL,GC-PN}
introduced a microscopic model of supercooled liquids, which is
based on three central ideas: (i) Particle mobility is sparse and
dynamics are spatially heterogeneous at times intermediate between
ballistic and diffusive motion. (ii) Particle mobility is the
result of dynamic facilitation, i.e., mobile particles assist
their neighbors to become mobile. (iii) Mobility propagation
carries a direction, the persistence length of which is larger for
fragile than for strong glass formers. Our present and
previous~\cite{MV-PRL} results support the main ideas of the
Garrahan-Chandler model on a qualitative level. Specifically, we
showed that SHD exists in BKS silica and, hence, this phenomenon
is not limited to non-strong model liquids. Moreover, it was
demonstrated in our previous work on BKS silica~\cite{MV-PRL} that
dynamic facilitation is important, which is further corroborated
by preliminary results for the Dzugutov liquid.~\cite{MB} Finally,
Garrahan and Chandler argue that the persistence length of the
direction of mobility propagation manifests itself in the
relevance of string-like motion so that the very limited
importance of this type of motion for BKS silica is consistent
with their idea of a shorter persistence length of particle flow
direction in strong liquids.

We conclude that, at least for BKS silica, there is a subtle
relation among dynamical processes on different time scales.
Specifically, we suggest that the different properties of viscous
liquids on the timescale of the structural relaxation are not only
a consequence of quantitative differences in SHD at intermediate
times, but also result from differences in the spatiotemporal
propagation of mobility. Consistently, we find that the evolution
of SHD at intermediate times is insensitive to the crossover in
the temperature dependence of the transport coefficients, and vice
versa.

Finally, we comment on the relevance of our findings with respect
to real silica. The BKS model is one of the simplest models of
silica, and neglects atomistic details such as three-body
interactions and charge transfer processes, which are included in
newer silica potentials.~\cite{JK1,JK2}  Both of these features
have been demonstrated to be important for, e.g., structural
transformations between silica crystal polymorphs.~\cite{JK2}
Despite its simplicity, the BKS model reproduces many bulk dynamic
and thermodynamic features of real silica. However, local dynamics
involving correlated or string-like motion in the melt may well be
sensitive to the re-distribution of charge during bond breakage
and to higher order terms in the interaction potential.
Investigation of SHD in more realistic silica models would
determine this, and are underway.

\section{Summary}

We showed that dynamics in BKS silica are spatially heterogeneous
and, hence, the structural relaxation in this model of a strong
liquid cannot be described as a statistical bond-breaking process,
as may have been expected from the Arrhenius-like temperature
dependence of the transport coefficients. Specifically, we
demonstrated that there are spatial correlations between mobile
and immobile particles, respectively, the extent of which grows
strongly upon cooling. In particular, the growth of dynamically
correlated regions continues in the Arrhenius temperature regime.
While the spatial correlations between highly mobile particles are
maximum in the late-$\beta$/ early-$\alpha$ relaxation regime of
the MCT, those between highly immobile particles are biggest at
later times close to the time constant of the structural
relaxation. On a qualitative level, all these findings for SHD in
BKS silica resemble that for non-strong model liquids.
Consistently, we found that defects of the local network structure
facilitate dynamics to some extent, but they are not a necessary
precondition for high particle mobility. On a quantitative level,
a detailed comparison with literature data showed that measures of
dynamic heterogeneity at intermediate times, e.g., the mean size
of clusters of mobile particles, exhibit a broad spectrum of
values at the lower end of which the present results for BKS
silica are found. Moreover, such comparison revealed that the
$\alpha$-relaxation is more nonexponential for liquids with
pronounced dynamic heterogeneity at intermediates times. However,
we found no evidence for a straightforward relation between the
properties of dynamics on different time scales so that the
correlations are subtle. In particular, the ``fragile-to-strong
crossover'' for the transport coefficients is not accompanied by a
substantial change in the temperature dependence of SHD at
intermediate times. On the other hand, our results are
qualitatively consistent with a microscopic model of viscous
liquids put forward by Garrahan and Chandler.~\cite{GC-PRL,GC-PN}
Following their ideas, we suggest that the spatiotemporal
characteristics of mobility propagation play an important role in
the differences between fragile and strong liquids.

\begin{acknowledgments}
We thank Y.\ Gebremichael and M.\ Bergroth for stimulating
discussions. M.\ V.\ gratefully acknowledges funding by the
Deutsche Forschungsgemeinschaft (DFG) through the Emmy
Noether-Programm.
\end{acknowledgments}

\end{document}